\documentclass[reqno,11pt]{article}
\usepackage{jheppub}
\usepackage{epsfig}
\usepackage{amssymb}
\usepackage{amsmath}
\usepackage{hyperref}
\usepackage{dsfont}
\usepackage{slashed}
\usepackage[dvipsnames]{xcolor}
\usepackage{epstopdf}
\usepackage{graphicx}

\def\p{\partial}
\def\l{\lambda}
\def\O{\Omega}
\def\L{\Lambda}
\def\S{\Sigma}
\def\s{\sigma}
\def\half{{1 \over 2}}
\def\a{\alpha}
\def\b{\beta}

\newcommand{\be}{\begin{eqnarray}}
\newcommand{\ee}{\end{eqnarray}}

\title{Elliptic  scattering equations}
\author[a]{Carlos Cardona}
\author[b,c]{and Humberto Gomez}
\affiliation[a]{Physics Division, National Center for Theoretical Sciences, National Tsing-Hua
University,\\ Hsinchu, Taiwan 30013, Republic of China.}
\affiliation[b]{Instituto de Fisica -- Universidade de S\~ao Paulo,\\
Caixa Postal  66318, 05315-970 S\~ao Paulo, SP, Brazil.}
\affiliation[c]{Facultad de Ciencias Basicas,  Universidad Santiago de Cali,\\
Calle 5 $N^\circ$  62-00 Barrio Pampalinda, Cali, Valle, Colombia.}

\emailAdd{carlosandres@mx.nthu.edu.tw}
\emailAdd{humgomzu@gmail.com}

\abstract{Recently the CHY approach has been extended  to one loop level using elliptic functions and modular forms over a Jacobian variety. Due to the difficulty in manipulating these kind of functions,  we propose  an alternative prescription that is totally algebraic. This new proposal is based on an elliptic algebraic curve embedded in a $\mathbb{C}P^2$ space. We show that for the simplest integrand, namely the ${\rm n-gon}$, our proposal indeed reproduces the expected result. By using the recently formulated $\Lambda-$algorithm, we found a novel recurrence relation  expansion in terms of tree level off-shell amplitudes. Our results connect nicely with recent results on the one-loop formulation of the scattering equations. In addition, 
this new proposal can be easily stretched out to hyperelliptic curves in order to compute higher genus.
}

\begin{document}

\begin{flushright}
\vspace{10pt} \hfill{NCTS-TH/1602} \vspace{20mm}
\end{flushright}

\maketitle


\section{Introduction}\label{sect:intro}

The tree-level S-matrix of massless particles in arbitrary dimensions can be written in an elegant form, reminiscent of string theory, as contour integrals over ${\cal M}_{0,n}$, the moduli space of $n$-punctured Riemann sphere \cite{Cachazo:2013hca}. In fact, some reformulations of field theory in terms of worldsheet amplitudes are nowadays understood from ambitwistor string theory \cite{Mason:2013sva, Casali:2015vta,Berkovits:2013xba,Gomez:2013wza}. Cachazo, He and Yuan (CHY) also extended their approach for the scattering of scalars, interactions among gauge bosons and gravitons and more others theories  \cite{Cachazo:2013hca,Cachazo:2013iea,Cachazo:2013iaa,Cachazo:2014xea,Cachazo:2014nsa}. 

The integrals proposed for these prescriptions give rise to rational functions of the kinematic invariants because they are localized to the solutions of a set of equations now known as the scattering equations. More precisely, if the location of the $a^{\rm th}$ puncture on the sphere is denoted by $\sigma_a$ and the momentum of the $a^{\rm th}$ particle is denoted by $k_a^\mu$ then the scattering equations are given by
\begin{equation}\label{CHYSE}
E_a=\sum_{b=1\atop b\neq a}^n \frac{k_a\cdot k_b}{\sigma_a-\sigma_b} =  0, \qquad a\in \{1,2,\ldots ,n\}.
\end{equation}
The scattering equations provide the link between the boundaries of the moduli space of $\mathbb{CP}^1$s with punctures to the factorizations regions of the space of kinematic invariants. It is this connection that lead to the search and construction of scattering amplitudes based on solutions to these equations. In other words, the solution space of the scattering equations corresponds to the natural environment where on-shell objects live. 

The CHY approach has been proved to produce the correct BCFW \cite{Britto:2005fq} recurrence relations in Yang-Mills and Cubic Scalar theories by Dolan and Goddard in \cite{Dolan:2013isa}.

Since the original formulation, many methods have been developed towards the use of CHY approach efficiently. In early attempts to deal with equations (\ref{CHYSE}), some solutions at particular kinematics were considered in \cite{Cachazo:2013iea, Kalousios:2013eca, Lam:2014tga} as well as at particular dimensions in \cite{Weinzierl:2014vwa, Cachazo:2013iaa, Cachazo:2016sdc, He:2016vfi}. Later, more general methods which avoid explicit finding of the solutions were developed \cite{Kalousios:2015fya,Dolan:2014ega, Huang:2015yka, Cardona:2015ouc, Cardona:2015eba, Dolan:2015iln, Sogaard:2015dba, Cachazo:2015nwa,mafra}. Parallel to those methods, generalized Feynman rules were developed \cite{Baadsgaard:2015ifa,Baadsgaard:2015voa} for 4-regular graphs containing single poles in the kinematic invariants. This has been generalized  very recently to the inclusion of higher poles in \cite{Huang:2016zzb}.

A natural further question is the generalization of scattering equations to Riemann surfaces of arbitrary genus and its connection to scattering amplitudes at higher orders in perturbation theory. This questions has been explored already in some directions. Starting with the construction of a  string theory in an ambitwistor space \cite{Mason:2013sva}, which reproduced the genus zero formula and provides a prescription for higher genus \cite{Adamo:2013tsa,Casali:2014hfa,Geyer:2015bja,Geyer:2015jch}. Later, upon previous construction of the scattering equations for massive particles \cite{Naculich:2014naa, Dolan:2013isa}, an extension of the CHY formalism at tree-level was made in which the loop momenta is emulated by taking a forward limit between two massive particles \cite{Cachazo:2015aol,He:2015yua}. This approach was recently generalized to planar higher loops in \cite{Feng:2016nrf}.

Nevertheless, despite all these remarkable developments, there are many issues that still need to be addressed. For example, how to obtain a handy prescription at genus $g=2$, or, how to perform in an efficient way CHY integrals at tree-level or even higher genus. We believe our work will prove to be useful to face these kind of difficulties.

As it was shown in \cite{Adamo:2013tsa},  one way to obtain the set of equations \eqref{CHYSE} is mapping a sphere with punctures to the null cone in a $D$ dimensional momentum space. This is done by introducing a Lorentz vector of meromorphic one-forms $\Omega^\mu$ on the sphere such that
\begin{equation}\label{general_Omega}
\frac{1}{2\pi i }\oint_{|z-\sigma_a|=\epsilon} \Omega^\mu = k_a^\mu \quad {\rm and} \quad \Omega^\mu\Omega_\mu = 0.
\end{equation}
In \cite{Adamo:2013tsa,Casali:2014hfa,Geyer:2015bja,Geyer:2015jch}, it was extended at one-loop level over a Jacobian variety and many interesting results have been obtained. However, although the prescription given in this work also generalized the Lorentz vector form, \eqref{general_Omega}, at genus $g=1$,  our approach is totally different  to the previous ones presented in \cite{Adamo:2013tsa,Casali:2014hfa,Geyer:2015bja,Geyer:2015jch}. In fact, our ideas follow the ones given in a very recent paper developed for one of the authors of this work \cite{Gomez:2016bmv}.

In \cite{Gomez:2016bmv},  a new reformulation for the tree-level scattering equations, which allow us to deal with off shell particles and higher poles in the kinematic invariants, has been presented.  In this paper we enhance that construction by considering genus one elliptic Riemann surfaces. Similarly as it was developed for the genus zero curve, the meromorphic differential $\Omega^\mu$ with the correct properties can be ensured by finding the solutions to a set of polynomial equations with coefficients which are rational in the kinematic invariants including the loop momenta. 
We will show that after applying a global residue theorem over the elliptic scattering prescription, one falls into the tree-level $\Lambda$ scattering approach of \cite{Gomez:2016bmv}. Later, one easily can perform the $\Lambda$  algorithm, which allows us to find in an elegant graphical way, a nice recurrence relation for the ${\rm n-gon}$ integrand. This recurrence relation can be written schematically as
\be\label{SchEx}
&{\cal I}_{n}(\ell)=
&\sum_{p=0}^{n-2}
{{\cal I}_{n-p-1}(\ell)\times{\cal I}_{p+1}(\ell)\over k_{1,2,\ldots p, \ell}}\, ,
\ee
which is very similar to the Q-cut discussed in \cite{Baadsgaard:2015twa,Huang:2015cwh}.

Let us do some remarks on our results to motivate the interested reader. The meromorphic differential written in terms of an elliptic curve bears a very simple form, which renders our method completely algebraic, without needing to deal with complicated Theta functions.  At the same time, our meromorphic differential has a straightforward generalization to higher genus curves.  In addition, after integrating over the modular parameter and particularize to the ${\rm n-gon}$ case, we recover previous results at one-loop level, which naturally appear as a consequence of our prescription, such as the rising of two extra massive particles at the forward limit which play the role of the loop-momenta.  
Our expansion (\ref{SchEx}) have some attractive properties. It provides a nice recurrence relation that allow us to write a $n-$particle amplitude in terms of lower point sub-amplitudes, similar to the Q-cut expansion. This lead us to think that our expansion might have applications to other theories at one loop as Yang-Mills or $\phi^4$, similarly as considered for the Q-cuts in \cite{Huang:2015cwh}. Even more, the technique we have used in this paper can be straightforwardly applied to other available integrands at one loop.

Finally, although we did not give an explicit proof for the ${\rm n-gon}$ conjecture formulated in \cite{Casali:2014hfa}, we believe our results provide the seed from where it should be easily proved.

The remainder of this paper is organized as follows. In section \ref{Sec2} we discuss some geometrical properties of algebraic curves describing Riemann surfaces of genus one that will become useful in our derivations.
In section \ref{Sec3} we present the meromorphic differential on a elliptic Riemann surface of genus one, from which we compute the associated scattering equations. Next in section \ref{SA_prescription} we present the prescription for the computation of the scattering amplitudes at one-loop.  By using the global residue theorem to perform the integration over the modular parameter of the torus in section \ref{Sec5}, we get a modified set of scattering equations that can be interpreted as off-shell tree-level CHY. In section \ref{Nodal} we show that the integration over the modulus leave us with two-spheres connected through a nodal point which simplifies the computation to a usual tree-level system. The given  tree-level system can be treated by using the $\Lambda-$algorithm in section \ref{Sec7} for general $n$ with particular lower particle examples discussed in section \ref{Sec8}. Finally, we make the discussion and conclusion on our results in section \ref{Sec9}.

\section{Elliptic Curve}\label{Sec2}

The genus $g$ of a Riemann surface given by a smooth plane curve of the degree $d$  embedded in $\mathbb{CP}^2$ can be computed by the formula
\be\label{genus}
g={(d-1)(d-2)\over 2}\,.
\ee

According to it, a genus-one surface corresponds to a curve of degree three. Thus,  any torus embedded in a $\mathbb{CP}^2$ with local coordinates $(z,y)$ can be described by the cubic curve
\be\label{cubiccurve}
y^2= z(z-\l_1)(z-\l_2)\,,
\ee
where $\l_1$ and $\l_2$ are complex parameters. Clearly, \eqref{cubiccurve} is invariant under the scale transformation
\begin{equation}\label{scaleS}
(y,z,\l_1,\l_2)  ~~ \rightarrow~~   (\kappa^3\, y,\kappa^2\,z,\kappa^2\,\l_1,\kappa^2\,\l_2), ~~{\rm where}~~ \kappa \in \mathbb{C}^*.
\end{equation}

The scale invariance in \eqref{scaleS} implies that  $(\l_1,\l_2)$ are the homogeneous coordinates  of a $\mathbb{C}P^1$ space, i.e. from the equivalence relation 
\begin{equation}
(\l_1,\l_2)\,\, \sim\,\, \kappa (\l_1,\l_2), ~~{\rm with}~~ \kappa \in \mathbb{C}^*,
\end{equation}
$(\l_1,\l_2)$  define a $\mathbb{C}P^1$.  We denote this equivalence class as 
$$
\langle\l_1,\l_2\rangle=\{ (\l_1,\l_2) \in \mathbb{C}^2-\{(0,0)\} : (\l_1,\l_2)\,\, \sim\,\, \rho (\l_1,\l_2),~~{\rm with}~~ \rho \in \mathbb{C}^*  \}.
$$
This projective space is  just the {\bf compact moduli space}\footnote{The Moduli space is defined as the space of conformally inequivalent curves.} of the elliptic curve \eqref{cubiccurve}.  Note that the point $(\l_1,\l_2)=(0,0)$ is excluded from the $\mathbb{C}P^1$ Moduli  space. This point is known as the cusp singularity and  it will not be included in our computations.  In fact, we are only interested  in the nodal singularities, which are related with the factorization limits, as we will see in section (\ref{Nodal}).

Following the same idea as in\cite{Gomez:2016bmv}, we define the holomorphic measure on this $\mathbb{C}P^1$  moduli space as
\begin{equation}\label{Mmoduli}
D\l:=\frac{\epsilon^{\a\b}\l_\a \, d\l_\b}{\l_1 \, \l_2\, (\l_1-\l_2)},~~{\rm with}~~ \a,\b=1,2.
\end{equation}
Note that this measure is not well defined on $\mathbb{C}P^1$ because it is not scale invariant. Nevertheless, this measure only makes sense into the elliptic scattering amplitude prescription, which will give in sectioin \ref{SA_prescription}.  
The factor, $\l_1 \, \l_2\, (\l_1-\l_2)$,  is just the square root of the discriminant of  the \eqref{cubiccurve} elliptic curve, i.e. 
 $\Delta[z(z-\l_1)(z-\l_2)]= \l_1^2 \, \l_2^2\, (\l_1-\l_2)^2$, and $\epsilon^{\a\b}\l_\a \, d\l_\b$ is the (1,0)-form invariant under the  $PSL(2,\mathbb{C})$ group. 

Finally, the global holomorphic form on the \eqref{cubiccurve} elliptic curve is given by \cite{Griffiths}
\begin{equation}
\omega =\rho \frac{dz}{y},
\end{equation}
where $\rho$ is a normalization constant such that\footnote{The ${\rm a-cycle}$ and ${\rm b-cycle}$ can be identified in Figure (1).} 
\begin{equation}\label{acyclecondition}
\oint_{\rm a-cycle} \omega(z) =1,
\end{equation}
i.e.
\begin{equation}
\frac{1}{\rho} = \oint_{\rm a-cycle} \frac{dz}{y}.
\end{equation}
The integration on the ${\rm b-cycle}$ is a function over $\langle\l_1,\l_2\rangle$
\begin{equation}
\oint_{\rm b-cycle} \omega(z) = f(\langle\l_1,\l_2\rangle).
\end{equation}
This  $f(\langle\l_1,\l_2\rangle)$ function  is know as the period matrix. In addition the global quadratic form is given  by
\begin{equation}
\omega ^2= \frac{dz\otimes dz}{y^2}.
\end{equation}

\section{Elliptic Scattering Equations.}\label{Sec3}
In this section we shall formulate the scattering equations over a Riemann surface of genus $g$ which admit a representation in terms of a elliptic curve. As it is well-known, all curves of genus $g=1$ admit an elliptic description.

\subsection{The meromorphic differential $\Omega^\mu$}

Let $\S_1$ be a Riemann surface of genus $g=1$ admitting a representation in terms of an elliptic curve,  such as in \eqref{cubiccurve}.

We would like to construct the most general meromorphic differential on $\S_1$ with only simple poles at $n$ points denoted by $(\s_a, y_a)$ with residue $k_a^\mu$, the particle momentum. The differential is given by
\be\label{omega}
\O^\mu =q^\mu\, {dz\over y} + \half\sum_{a=1}^n \left({y_a\over y}+1 \right) {k_a^\mu \,dz\over z-\s_a}~ ,
\ee
on the support of the curves
\begin{equation}
y^2=z(z-\l_1)(z-\l_2),~~ y_a^2=\s_a(\s_a-\l_1)(\s_a-\l_2),~~ {\rm with}~~ a=1,\ldots n.
\end{equation}

The factor $\left({y_a\over y}+1 \right)$ is there to ensure that the pole $1/(z-\s_a)$ is located on the same branch as the puncture $(\s_a, y_a)$. The first terms in \eqref{omega}, $q^\mu\, dz/ y$,   parametrizes the freedom one has in adding any holomorphic differential. 

\subsection{Scattering Equations}\label{E1_a}

Having constructed the momentum differential $\O^\mu$ we can proceed to imposing the massless condition $\O^2=\O^\mu\O_\mu=0$. This is the condition that links the moduli space of genus $g=1$ Riemannn surfaces  with $n$ marked points to the space of kinematic invariants with coordinates $s_{ab}=k_a\cdot k_b$ (subject to constraints from momentum conservation and the on-shell condition $k^2_a=0$).

Expanding $\O^2$  around $z=\s_a$ and $y=y(\s_a)$, where $y^2(\s_a)=y^2_a=\s_a(\s_a-\l_1)(\s_a-\l_2)$, 
can be on any branch, one finds
$$
\left({y_a\over y(\s_a)}+1 \right)\left({ q\cdot k_a \over y(\s_a)} +\half\sum_{b=1\atop b\ne a}^n \left({y_b\over y(\s_a)}+1 \right) {k_a\cdot k_b \over \s_a-\s_b} \right){dz\otimes dz \over z-\s_a}.
$$
One has to require that this vanishes both when $y(\s_a) = y_a=$ and when $y(\s_a) = - y_a$. Clearly, the latter is trivially satisfied due to the presence of the prefactor. This means that the only
equations we have to impose is the vanishing the second factor when $y(\s_a) = y_a$, i.e.,
\be\label{esa}
 E^1_a:={q\cdot k_a \over y_a}+ \half\sum_{b=1\atop b\ne a}^n \left({y_b\over y_a}+1 \right) {k_a\cdot k_b \over \s_a-\s_b}=0,
\qquad a\in \{1,2,...,n \}.
\ee
We call these equations the {\bf elliptic scattering equations} and they are the  genus $g=1$ generalization of the tree level scattering equations given by \cite{Gomez:2016bmv}
\begin{equation}
 E^T_a:=\half\sum_{b=1\atop b\ne a}^n \left({y^T_b\over y^T_a}+1 \right) {k_a\cdot k_b \over \s_a-\s_b}=0,
\quad {\rm where } ~~ (y^T_a)^2=\s^2_a-\L^2.  
\end{equation}

The $n$ equations in \eqref{esa} are clearly necessary to ensure that $\O^2=0$ but they are not sufficient.  Let us note that  the meromorphic form in \eqref{omega} is a (1,0) global form on a elliptic curve with $n$ marked points $\{\s_1,\ldots,\s_n\}$, i.e. a torus with $n$ punctures at positions $\s_a$. The Moduli space of this surface, which we call ${\cal M}_{1,n}$,  has complex dimension ${\rm dim}_{\mathbb{C}}({\cal M}_{1,n})=n$. However, the $n$ elliptic scattering equations $E^1_a$
are not linearly independent, which can be  inferred from the identity
\begin{equation}\label{symmetry}
\sum_{a=1}^n\,y_a\,E^1_a=0,
\end{equation}
where only momentum conservation has been used\footnote{The identity \eqref{symmetry} is a consequence of a global symmetry on the elliptic curve, or in other words, it due to the existence of a global vector field  on the curve.}. Hence, we must impose one more constraint so as to guarantee $\Omega^2=0$.

Let us remember  that the elliptic scattering equations \eqref{esa} were  obtained by expanding the $\Omega^2$ quadratic form around each $(\s_a,y_a)$ puncture, so  the only thing one can say is that on the support of these equations the $\Omega^2$ form is a global quadratic form on the elliptic curve without punctures, i.e.
\begin{equation}
\Omega^2\Big|_{E^1_a=0}=\frac{\cal L}{y^2}\,\, dz\otimes dz,
\end{equation}
where ${\cal L}$ is a constant over $z$. Thus, in order  to ensure that $\Omega^2$ vanishes we must impose the constraint 
\begin{equation}
{\cal L}=0.
\end{equation}
From the $\Omega^\mu$ form  in \eqref{omega} it is straightforward to see  
\begin{equation}
\Omega^2 = \,\frac{{\cal L}}{y^2}\, dz\otimes dz, 
\end{equation}
where 
\begin{equation}\label{Aomega2}
{\cal L}=\left[q^\mu + \frac{1}{2}\sum_{a=1}^n \frac{k_a^\mu}{z-\s_a}(y_a+y)\right]^2\,\, .
\end{equation}
Note that the (1,0)-form given by
$$
\Omega^2(z)\,\left(\rho {y\over dz}\right),
$$
must be  proportional to the global holomorphic form, $\omega=\rho \,dz/y$,
on the support of the elliptic scattering equations, $E^1_a=0$.  Therefore,  instead of work with the expression found in \eqref{Aomega2} we can use the property \eqref{acyclecondition} and so we define the ${\cal L}$ constraint  as
\begin{equation}\label{Adefinition}
{\cal L}:=\rho  \oint_{\rm a-cycle}\Omega^2(z)\,\left({y\over dz}\right)=\rho 
\oint_{\rm a-cycle} \left[q^\mu + \frac{1}{2}\sum_{a=1}^n \frac{k_a^\mu}{z-\s_a}(y_a+y)\right]^2{dz \over y}=0,
\end{equation}
where we have  chosen the ${\rm a-cycle}$ on the upper branch, i.e. $y=\sqrt{z(z-\l_1)(z-\l_2)}$.

Finally, we have found the whole set of constraints that ensure the vanishing of the $\Omega^2(z)$ quadratic form 
\begin{equation}
E^1_a=0 , ~~{\cal L}=0,~~{\rm with} ~~  a=1,\ldots, n.
\end{equation}

\subsection{Global Vector Field}\label{gvf}

Genus one Riemann surfaces are also special. It is well known that if a torus with punctures is described by its Jacobian variety, then for fixed $\tau$, the punctures can be all simultaneously translated by the same amount without changing the complex structure.
This means that one of the $n$ puncture locations can be fixed to a particular value on the elliptic curve. This invariance is manifested itself as a linear dependence among the elliptic scattering equations, as it was shown in \eqref{symmetry}.  

It is interesting to understand the source of this redundancy. A straightforward translation on the Jacobian variety is given by  the global holomorphic vector field  $V=\p_x$. Naively, this vector field is mapped on the elliptic curve to the vector  ${\cal V}=y\p_z$, nevertheless this is not a viable possibility as the former was defined for fixed $\tau$ while the latter can change $[\l_1,\l_2]$ on the support of the elliptic scattering equations. One can verify that the  holomorphic vector field 
\begin{equation}\label{vecgenusone}
{\cal V}=\sum_{a=1}^n  y_a\,\p_{\s_a}+{1\over 4}\left(\sum_{a=1}^n \s_a\,k_a^\mu\right) \,\p_{q^\mu} \,,
\end{equation}
is the generator of  the symmetry in the elliptic scattering equations on the support of the elliptic curve.

Fixing one puncture location, for example $\s_i$,  the Fadeev-Popov determinant coming from this gauge fixing is given by
\begin{equation}\label{FP}
\Delta_{\rm FP}(i)=y_i.
\end{equation}

\section{Scattering Amplitude Prescription}\label{SA_prescription}

Following the CHY prescription \cite{Cachazo:2013hca} along with \cite{Gomez:2016bmv}, let us propose the following S-matrix
\begin{equation}\label{HyperSmatrix}
A_n=\frac{1}{{\rm Vol}(G)} \int d^Dq\wedge \int_\Gamma {D\l \over {\cal L}}\wedge \left(\prod_{a=1}^{n}\, {dy_a \over  C_a}\right)\wedge \left(\prod_{b=1}^{n} \, {d\s_b \over y_b} \right) {H(\s,y) \over \prod_{c=1}^n E^1_c} \,,
\end{equation}
where $D\l$ is the measure over the tori Moduli space given in \eqref{Mmoduli}, $G$ is the gauge group generated by the ${\cal V}$ global vector field given in \eqref{vecgenusone} and $C_a$'s  are 
\begin{equation}\label{Ccurves1}
C_a=y_a^2-\s_a(\s_a-\l_1)(\s_a-\l_2).
\end{equation}

The $A_n$  integral can be justified as follows. The $d y_a/C_a$'s integrals  are given to support the prescription on the elliptic curves 
\begin{equation}\label{Ccurves2}
C_a=0, ~~{\rm where }~~ a=1,\ldots, n,
\end{equation}
i.e. one can say these constraints define the integration contours over the $y_a$'s variables.  The $d\s_a/y_a$ factor is the only one holomorphic form on the elliptic curve $C_a=0$. The denominator, ${\cal L}\,\prod_b  E^1_b$, is just the product of the  elliptic scattering equations, i.e the constraints
\begin{equation}
E^1_b=0,~~~{\cal L}=0,~~{\rm with }~~b=1,\ldots,n
\end{equation}
define the integration contours over the $\s_i$'s variables and the $[\l_1,\l_2]$ coordinate over the tori Moduli space.  Therefore, the total integration contour, $\Gamma$,  is defined by the equations
\begin{equation}
C_a=0, ~~E^1_a=0,~~~{\cal L}=0,~~{\rm with }~~a=1,\ldots,n
\end{equation}
The $H(\s,y)$ function is the integrand which defines a theory. Finally, the $d^Dq$ measure  is the integration over the freedom to add a global holomorphic form in $\Omega^\mu(z)$ given in \eqref{omega}. The integration contour over the $q^\mu$ variables is not specified yet. 

Nevertheless, although we have justified the \eqref{HyperSmatrix} integral, we need to check that it is in fact a well defined prescription on  ${\cal M}_{1,n}$ Moduli space by showing that the holomorphic top form given by
\begin{equation}
\Phi:=\phi(\s,y,\l_\a) H(\s,y),
\end{equation}
where
\begin{equation}
\phi(\s,y,\l_\a):= d^Dq\wedge {D\l \over {\cal L}} \wedge \left(\prod_{a=1}^{n}\, {dy_a \over C_a}\right)\wedge \left(\prod_{b=1}^n {d\s_b \over E^1_b}\right)  {1 \over \prod_{c=1}^n y_c} ,
\end{equation}
is invariant by the global holomorphic vector field ${\cal V}$.  This means the Lie derivative  
\begin{equation}
L_{\cal V} (\Phi) = \left[ L_{\cal V} (\phi)\right] H(\s,y) +  \phi(\s,y,\l_\a) \,{\cal V}(H(\s,y))
\end{equation}
must vanish on the support of the elliptic scattering equations. It is straightforward to check  
$$
L_{\cal V} (\phi)=0,
$$
hence in order to vanish $L_{\cal V} (\Phi) $ we must require the condition ${\cal V}(H(\s,y))=0$.

Let us consider the particular case when  $H(\s,y)=1$. Clearly the condition ${\cal V}(H(\s,y))={\cal V}(1)=0$ is trivially satisfied. So, the integral 
\begin{equation}\label{ngon}
{\rm A}_n^{\rm n-gon}(1,\ldots,n):= \frac{1}{{\rm Vol}(G)} \int d^Dq\wedge \int_\Gamma {D\l \over {\cal L}}\wedge \left(\prod_{a=1}^{n}\, {dy_a \over  C_a}\right)\wedge \left(\prod_{b=1}^{n} \, {d\s_b \over E^1_b} \right) {1 \over \prod_{c=1}^n y_c} 
\end{equation}
is well defined on ${\cal M}_{1,n}$ and it is know as the ${\bf n-gon}$.

In the rest of the paper we will work just with the ${\rm n-gon}$ integral.

\section{Gauge Fixing and the Global Residue Theorem}\label{Sec5}

To gauge the freedom coming from the invariance generated by (\ref{vecgenusone}) we fix the coordinate $\s_n$ and drop the scattering equation $E_n^1$. Thus, the ${\rm A}_n^{\rm n-gon}$ integral becomes
\begin{equation}\label{ngonG}
{\rm A}_n^{\rm n-gon}(1,\ldots,n)=  \int d^Dq\wedge \int_\Gamma {D\l \over {\cal L}}\wedge \left(\prod_{a=1}^{n}\, {dy_a \over  C_a}\right)\wedge \left(\prod_{i=1}^{n-1} \, {d\s_i \over E^1_i} \right) {\Delta^2_{\rm FP}(n) \over \prod_{b=1}^n y_b} \Big|_{\s_n={\rm cte}}
\end{equation}
where  we have introduced two Faddeev-Popov determinants $\Delta_{\rm FP}$, one for  fixing  $\s_n$ and the other to gauge the $E^1_n$ scattering equation. The $\Gamma$ contour is defined by the solution of the $2n$ equations
\begin{equation}\label{contour}
{\cal L}=0, ~~ C_a=0, ~\,\, a=1,\ldots,n, ~~ E_i^1=0, ~\,\, i=1,\ldots, n-1.
\end{equation}
Note that the $\s_n$ position can be gauged at any point on the curve, except at the branch points.

The ${\rm A}_n^{\rm n-gon}$ integral is not a simple computation, in fact, solving the equations given in \eqref{contour} is a very hard task in general. Nevertheless, in a similar way as it was done in  \cite{Geyer:2015bja} and \cite{Gomez:2016bmv},  we can apply the global residue theorem over the $\mathbb{C}P^1$ Moduli space  of the elliptic curve, i.e over the $\langle\l_1,\l_2\rangle$ coordinate (see section \ref{Sec2}),  with a view to simplify the computation.  To perform this residue theorem we choose the chart $U_1= \{ \langle\l_1,\l_2\rangle =(1, \l) : \l\in \mathbb{C}\}$ on the $\mathbb{C}P^1$ Moduli space, thus the $D\l$ measure and the $C_i$ contours become
\begin{equation}
D\l \Big|_{U_1}= \frac{d\l}{\l (1-\l)},\qquad  C_a=y_a^2-\s_a(\s_a-1)(\s_i-\l),~~{\rm where } ~ a=1,\ldots n.
\end{equation}

In order to perform  a residue theorem over $\l$, we perceive that the only  dependence over this variable in ${\rm A}_n^{\rm n-gon}$ is given by the denominator $\l(1-\l)\prod_{a}C_a$. Note also that the ${\cal L}=0$ constraint only depends over the $\s_i$'s and $y_a$'s variables. Thus, it is enough to write the following piece of the  ${\rm A}_n^{\rm n-gon}$ integrand
\begin{equation}
\frac{d\l}{\l (1-\l)}\times \frac{\prod_{a=1}^n  dy_a}{\left(\prod_{a=1}^n C_a\right) \, {\cal L}}\ldots  = \frac{1}{\l (1-\l)}\times\left(\frac{d\l}{C_j}\right)\times \prod_{a\neq j}^n\left(    \frac{  dy_a}{ C_a}\right) \times \left(\frac{dy_j}{{\cal L}}\right)\ldots ,
\end{equation}
where we  say that the denominator, $\left(\prod_{a=1}^n C_a\right){\cal L}$, defines the integration contour  for the $n+1$ variables, $(\l,y_1,\ldots,y_n)$.  Without loss of generality,  we choose the denominator $C_j$ to fix the integration contour over $\l$. Hence, applying a residue theorem over it one obtains
\begin{equation}
\frac{1}{\l (1-\l)}\times\left(\frac{d\l}{C_j}\right)\times \prod_{a\neq j}^n\left(    \frac{  dy_a}{ C_a}\right) \times \left(\frac{dy_j}{{\cal L}}\right)\ldots=
\frac{-1}{C_j}\times\left(\frac{d\l}{\l (1-\l)}\right)\times \prod_{a\neq j}^n\left(    \frac{  dy_a}{ C_a}\right) \times \left(\frac{dy_j}{{\cal L}}\right)\ldots, \nonumber
\end{equation}
where the $C_j$ contour has been changed by the new one $\l (1-\l)=0$ and $C_j$ becomes part of the integrand.

So as to  recover the $C_j$  constraint we must again perform a global  residue theorem, but now over $y_j$. Before computing this global residue theorem we must  rewrite the denominator in ${\rm A}_n^{\rm n-gon}$ as a polynomial over the $y_j$ variable. So, the integration  by $y_j$ into ${\rm A}_n^{\rm n-gon}$ can be rewritten as
\begin{equation}
\frac{1}{y_j\,\,C_j}\times
\left(\frac{dy_j}{\cal L}\right)\times \left(\prod_{i=1}^{n-1}\frac{d\s_i}{E_i^1}\right)\ldots=
\frac{\prod_{a\neq j}^n y_a}{C_j}\times
\left(\frac{dy_j}{\cal L}\right)\times \left(\prod_{i=1}^{n-1}\frac{d\s_i}{\tilde E_i^1}\right)\ldots,
\end{equation}
where we have defined 
\begin{equation}
\tilde E^1_a= q\cdot k_a + \frac{1}{2}\sum_{b\neq a}\frac{k_a\cdot k_b}{\s_{ab}}(y_a + y_b).
\end{equation}
Clearly,  the denominator, ${\cal L}\,\,\prod_{i=1}^{n-1}\tilde E_i^1$, which defines the integration contour over the $n$ variables, $(y_j,\s_1,\ldots,\s_{n-1})$, is a polynomial over $y_j$. Without loss of generality,  we can say that the ${\cal L}$ factor fixes the integration contour over $y_j$, so,
using a global residue theorem over $y_j$ we obtain
\begin{equation}
\frac{\prod_{a\neq j}^n y_a}{C_j}\times
\left(\frac{dy_j}{\cal L}\right)\times \left(\prod_{i=1}^{n-1}\frac{d\s_i}{\tilde E_i^1}\right)\ldots=
-\frac{\prod_{a\neq j}^n y_a}{\cal L}\times
\left(\frac{dy_j}{C_j}\right)\times \left(\prod_{i=1}^{n-1}\frac{d\s_i}{\tilde E_i^1}\right)\ldots,
\end{equation}
where the ${\cal L}$ contour has been changed by the new one $C_j=0$ and ${\cal L}$ becomes part of the integrand.

Finally, after performing these two residues theorem the ${\rm A}_n^{\rm n-gon}$ integral can be read  as
\begin{equation}\label{ngonG}
{\rm A}_n^{\rm n-gon}(1,\ldots,n) =  \int d^Dq\wedge \int_\gamma \frac{d\l}{\l(1-\l)}\wedge \left(\prod_{a=1}^{n}\, {dy_a \over C_a}\right)\wedge \left(\prod_{i=1}^{n-1} \, {d\s_i \over E^1_i}\right)  \left( {\Delta^2_{\rm FP}(n) \over {\cal L}\,\prod_{b=1}^n y_b} \right)\Big|_{\s_n={\rm cte}},
\end{equation}
where $\s_n$ is a constant\footnote{$\s_n$ is a constant  such that $\s_n\neq 0,1,\infty$. Note that $\{0,1,\infty\}$ are the branch points. }
and  the new contourn, $\gamma$,  is defined by the equations
\begin{equation}
\l(1-\l)=0,~~    C_a=0, ~a=1,\ldots,n,~~  E_i^1=0,~ i=1,\ldots, n-1.
\end{equation}

In section \ref{Nodal} and \ref{Sec7} we will show that using this new integration contour the ${\rm A}_n^{\rm n-gon}$ integral is trivially solved.

\section{Nodal Singularities}\label{Nodal}

The idea of this section is to compute the  integration over the $\mathbb{C}P^1$ Moduli space, i.e over the $\langle\l_1,\l_2\rangle$ variable.

As it was shown in the previous section, the integration over $\langle\l_1,\l_2\rangle$ on the $U_1= \{ \langle\l_1,\l_2\rangle =(1, \l) : \l\in \mathbb{C}\}$ chart and the $y$'s variables is given by 
\begin{equation}
\frac{d\l}{\l(1-\l)} \wedge \prod_{i=1}^n\frac{dy_i}{C_i}\ldots\,\, ,
\end{equation}
where the contour is defined by the equations
\begin{equation}
\l(1-\l)=0,~~ C_i=0, ~~i=1,\ldots n .
\end{equation}
The equation $\l(1-\l)=0$ defines the contour over $\l$  and $C_i=0$ defines the contour over $y_i$ for each $i=1,\ldots n$.

Clearly, the integration over $\l$ implies that  one must evaluate the integrad into ${\rm A}_n^{\rm n-gon}$  at $\l=0$ and $\l=1$.  Nevertheless,  in order to explore the whole $\mathbb{C}P^1$ Moduli space we now consider the chart
$$
U_2= \{ \langle\l_1,\l_2\rangle   =(\tilde \l, 1) : \tilde\l\in \mathbb{C}\}.
$$
On this chart the $D\l$ measure and  the $C_i$'s  constraints keep the same form
 \begin{equation}
D\l\Big|_{U_2}=\frac{d\tilde\l}{\tilde\l (1-\tilde\l)}, \qquad ~ C_i=y_i^2-\s_i(\s_i-\tilde\l)(\s_i-1)=0,\quad  i=1,\ldots, n .
\end{equation}  
Note that the integration over $\tilde \l$ at the point $\tilde \l=1$  has been already computed when one makes $\l=1$ on  $U_1$. So, there is  only a point which  must be evaluated on $U_2$, this point is at $\tilde \l=0$, i.e.  $\l=\infty$.  Finally, we have obtained that the integration over $\l$ means that one must evaluate the integrand
of ${\rm A}_{n}^{\rm n-gon}$  at the three branch points on the elliptic curve, i.e. at $\l=\{0,1,\infty\}$. They are the points where the curve becomes a degenerate torus, such as it is shown in figure (1)
\begin{center}
\includegraphics[scale=0.6]{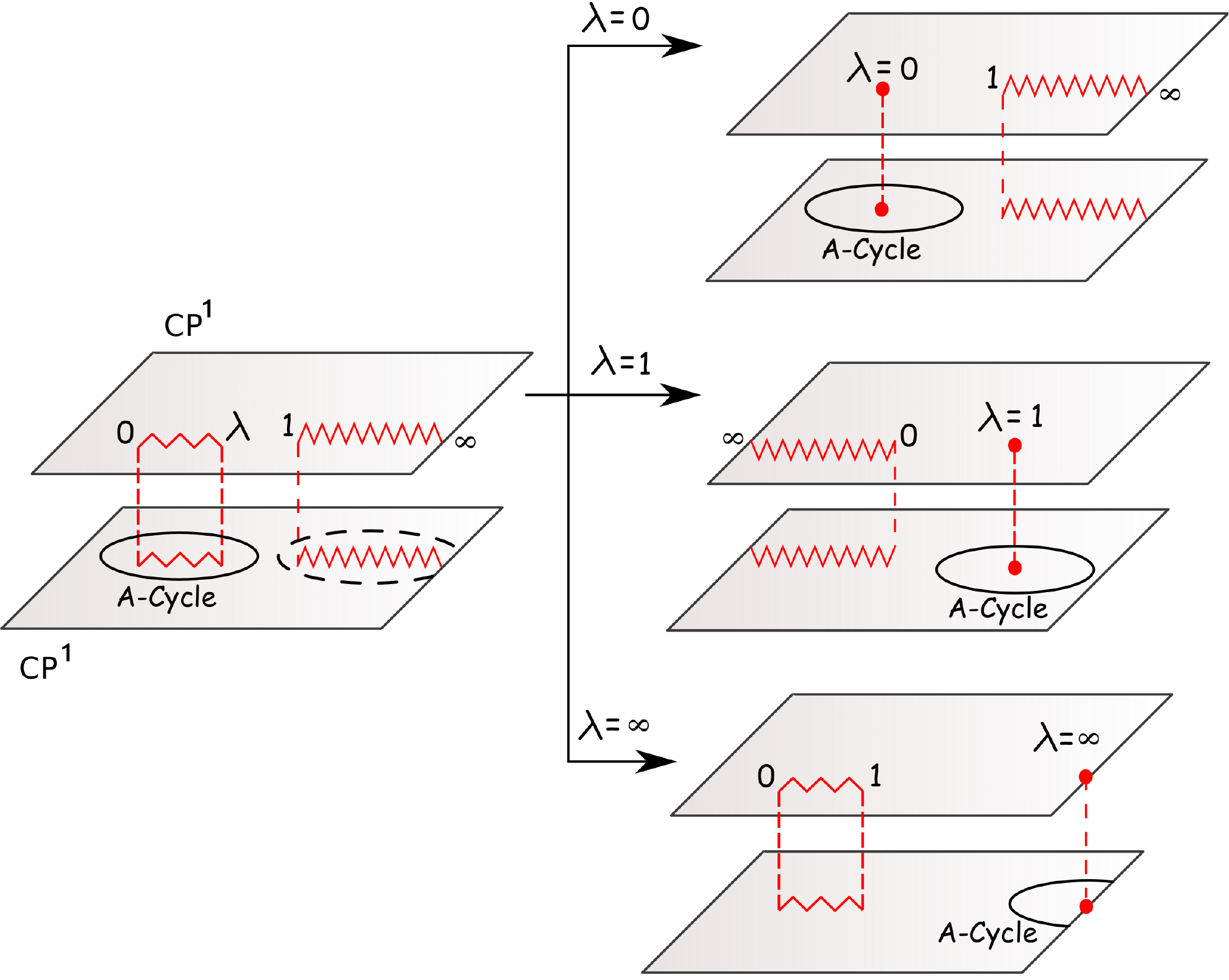}
\begin{center}
Figure 1: \,{\small {\rm Nodal singularities of the Elliptic curve\, .}}
\end{center}
\end{center}
These three singularities are known as the nodal singularities  and  as it can be noted in figure (1), they represent the same degenerates torus (pinched torus), therefore, we will only concentrate  on one of them, $\l=0$.

\subsection{Scattering Amplitude at $\l=0$.}\label{factorization}

In this section we compute the ${\rm A}_{n}^{\rm n-gon}$ integral  at $\l=0$.

First of all, it is straightforward to  see that at the point $\lambda=0$  the $C_i$ contour, i.e. the elliptic  curve, simplifies to
\begin{equation}\label{Ctree}
C_i=y_i^2-\s_i^2(\s_i-1):=\s_i^2[(y^{\rm T}_i)^2-(\s_i-1)]:=\s_i^2 C^{\rm T}_i\,,
\end{equation}
where we have defined both $y_i:=\s_i\,y^{\rm T}_i$ and $C^{\rm T}_i:=(y^{\rm T}_i)^2-(\s_i-1)$. By looking at the genus formula (\ref{genus}) we see that due $C^{\rm T}_i$ has degree one it corresponds to a sphere and so its why we have used the superscript ${\rm T}$ referring  to ``Tree". As we can interpret from figure (1),   the curve, $\s^2[(y^{\rm T})^2-(\s-1)]=0$, 
is just a degenerated  torus into two-spheres connected by a fixed branch cut and two new  punctures  arise, which are  distributed on both sheets at  $\s_{n+1}=\s_{n+2}=0$.  

The elliptic scattering equations at $\lambda=0$, i.e. using the transformation, $y_a=\s_a\,y_a^{\rm T}$, become
\begin{equation}\label{lamzerores}
 E_a^1\Big|_{\l=0}= {({\rm I} \ell)\cdot k_a \over \s_a y^{\rm T}_a}+ \half\sum_{b=1\atop b\ne a}^n \left({y^{\rm T}_b \over y^{\rm T}_a}+1 \right) {k_a\cdot k_b \over \s_{ab}} \,.
\end{equation}
where ${\rm I}=\sqrt{-1}$ and we have defined
\begin{equation}
{\rm I} \ell^\mu := q^\mu-\frac{1}{2}\sum_{b=1}^n y_b^{\rm T}\, k^\mu_b.
\end{equation}
It is straightforward to note that the elliptic scattering equations in \eqref{lamzerores} can be written as
\begin{equation}\label{treeSE}
E_a^1\Big|_{\l=0}= \half\sum_{A=1\atop A\ne a}^{n+2} \left({y^{\rm T}_A\over y^{\rm T}_a}+1 \right) {k_A\cdot k_a \over \s_{Aa}}:= E^{\rm T}_a\,, ~~ {\rm with}~~ a=1,\ldots n,
\end{equation}
where \footnote{Please notice that $\s_{n+1},\,\s_{n+2}$ were called $\s_{\ell^+},\,\s_{\ell^-}$ and $\s_{+},\,\s_{-}$  in \cite{Geyer:2015jch} and in \cite{Cachazo:2015aol} respectively.}
\begin{align}\label{Sn+1}
& \s_{n+1}=0, ~~ y_{n+1}^{\rm T}=\sqrt{\s_{n+1}-1}=\sqrt{-1}={\rm I},\nonumber\\
& \s_{n+2}=0, ~~ y_{n+2}^{\rm T}=-\sqrt{\s_{n+1}-1}=-\sqrt{-1}=-{\rm I},\\
& k_{n+1}^\mu=\ell^\mu, ~~ k_{n+2}^\mu=-\ell^\mu\nonumber .
\end{align}
So, the elliptic scattering equations at $\l=0$,  $E^{1}_a\Big|_{\l=0}:=E^{\rm T}_a$, are  just the tree level scattering equations  for $n+2$ particles given in the $\L$ prescription formulated in \cite{Gomez:2016bmv}.  The two extra particles  or punctures, which are  located on different branches of the double cover, have opposite momentum, i.e. $k_{n+1}^\mu=-k_{n+2}^\mu:=\ell^\mu$,  but, what is the physical meaning of $\ell^\mu$? In order to solve this question we compute the integral over the ${\rm a-cycle}$  (see figure (1)) of the $\Omega ^\mu(z)$ (1,0)-meromorphic form given in (\ref{omega})
\begin{equation}\label{Flux}
\oint_{\rm a-cycle}\Omega^\mu(z)\Big|_{\l=0}=\oint_{|z|=\epsilon}\left[q^\mu+\half \sum_{a=1}^n\frac{k_a^\mu}{z-\s_a}(y_a+y)
\right]_{\l=0}\frac{dz}{y}=-{\rm I} ( q^\mu -\half \sum_{a=1}^n y_a^{\rm T} k_a^\mu )
\end{equation}
where we have considered the ${\rm a-cycle}$ on the upper sheet, i.e. $ y=z\,\sqrt{z-1}$, and the support  on $C_a=0$, i.e. $y_a=\s_a\,y_a^{\rm T}$. Hence,  from \eqref{Flux} one can conclude that $\ell^\mu$ is just the flux of the $\Omega^\mu(z)$ meromorphic form  around the ${\rm a-cycle}$
\be\label{lmu}
\ell^\mu=\oint_{\rm a-cycle}\Omega^\mu(z)\Big|_{\l=0}.
\ee
It is natural to identify the {\bf flux momentum}, $\ell^\mu$, as the loop momentum of a one-loop graph. Additionally, we can immediately see from equation  (\ref{Flux}) that the condition $k_{n+1}^\mu=-k_{n+2}^\mu:=\ell^\mu$ arise naturally as long as the punctures $(n+1)$ and $(n+2)$ are sit on different sheets, because the relative sign in the $y-$coordinate for different sheets induce a relative sign in (\ref{Flux}).

With (\ref{lmu}) in mind, it is simple to carry out the ${\rm a-cycle}$ integral in the ${\cal L}$ definition
\begin{equation}
{\cal L}\Big|_{\l=0}=\rho \oint _{|z|=\epsilon}\left[q^\mu+\half \sum_{a=1}^n\frac{k_a^\mu}{z-\s_a}(y_a+y)
\right]^2_{\l=0}\frac{dz}{y}=-\ell ^2,
\end{equation}
where we have used $\rho|_{\l=0}={\rm I}$.
Since ${\cal L}$ is not anymore a scattering equation in ${\rm A}_{n}^{\rm n-gon}$, i.e. ${\cal L}\neq 0$,  then the {\bf flux momentum}, $\ell^\mu$, is off-shell, $\ell^2\neq 0$,  in addition, it is worth noting that the momentum conservation for the $n+2$ particles is still satisfied
$$
\sum_{A=1}^{n+2}k_A^\mu = k_1^\mu+\ldots + k_n^\mu+\ell^\mu + (-\ell^\mu) =0.
$$

Finally, the measure over the $y_a$'s variables become
\begin{equation}
\left(\prod_{a=1}^n \frac{d y_a}{C_a}\right) \frac{y_n^2}{\prod_{b=1}^n y_b} = 
\left(\prod_{a=1}^n \frac{d y^{\rm T}_a}{C^{\rm T}_a}\right) \frac{(\s_n y^{\rm T}_n)^2}{\prod_{b=1}^n \s_b^2\, y^{\rm T}_b},
\end{equation}
and therefore ${\rm A}^{\rm n-gon}_n$  can be read as
\begin{equation}\label{HyperSmatrixForward}
{\rm A}_n^{\rm n-gon}(1,\ldots,n) =-{\rm I}^{D}\int    {d^D\ell  \over\ell^2} \int_{\gamma} \left( \prod_{a=1}^n
{ y^{\rm T}_a\,d y^{\rm T}_a \over C^{\rm T}_a}\right)
\left(
\prod_{i=1}^{n-1}\,{d\s_i \over E^{ \rm T}_i}\right)
 { (\s_n\,y^{\rm T}_n)^2\over   \prod_{b=1}^n ( \s_b\,y^{\rm T}_b)^2}\,,
\end{equation}
where $\s_n$ is fixed such that $\s_n\neq 0,1,\infty$ and the  $\gamma$ contour is defined by the $2n-1$ equations
\begin{equation}
C_a^{\rm T}=0,~~a=1,\ldots, n, ~~~E_i^{\rm T}=0,~~ i=1,\ldots, n-1.
\end{equation}

\subsection{${\rm A}^{\rm n-gon}_n$ and the $\L$ prescription}\label{Agon}

So far, we have found that the elliptic scattering equations at $\l=0$ become the tree level scattering equations. Now, in order to clarify the meaning of the S-matrix integrand in (\ref{HyperSmatrixForward}),  we show in this section that in fact (\ref{HyperSmatrixForward}) is  a tree level expression written in terms of the $\L$ prescription given in \cite{Gomez:2016bmv}.

Let us introduce the third-kind form for the quadratic curve, $(y_a^{\rm T})^2=\s_a-1$, as
\begin{equation}\label{tau}
\tau_{a:b}:={1\over 2\,y^{\rm T}_a}\left({y^{\rm T}_a+y^{\rm T}_b\over\s_{ab}}\right)={1\over 2\,y^{\rm T}_a}\left({1 \over y^{\rm T}_a-y^{\rm T}_b}\right)\,.
\end{equation}
The motivation for the above definition is the following identity  (on the upper branch)
\begin{equation}\label{mapping}
\tau_{a:b}\,d\s_a={d z_a\over z_{ab}},\,~ ~~ {\rm with ~the ~transformation} ~~~ \s_a=z_a^2+1,
\end{equation}
where the $z_a$'s variables are the usual coordinates over the sphere in the original CHY approach. So, the \eqref{mapping} transformation gives us the map to the original CHY integrals. In addition, notice that the $\tau_{a:b}$ form and the transformation defined in \eqref{tau} are simpler to the ones given in \cite{Gomez:2016bmv}, the reason is because in \cite{Gomez:2016bmv} the quadratic curve is a little more complicated, $y_a^2=\s_a^2-\L^2$.

As it was shown in \cite{Gomez:2016bmv},  making chains of $\tau$'s translate in the usual chains of $1/(z_{ab})$  factors in the CHY formalism. However, since we have fixed $\s_{n+1}$ and $\s_{n+2}$ on top of each other in different branches, we should be careful with chains involving them. 

Let us consider the following product
\begin{equation}\label{tauprod}
\tau_{a:n+1}\tau_{n+2:a}=-\tau_{a:n+2}\tau_{n+1:a}={1\over 2^2\,{\rm I}\,\,y^{\rm T}_a\,\s_a}\, ,
\end{equation}
where we have used \eqref{Sn+1}.
This product allow us to rewrite one of the factors in (\ref{HyperSmatrixForward}) as,
\begin{equation}\label{chainfactor}
\prod_{a}^n{1\over (\s_a\,y^{\rm T}_a)^2}=2^{4n}\prod_{a=1}^n (a:n+1)(a:n+2)\,\, ,
\end{equation}
where we have used the chain notatation
\begin{equation}
(i_1: i_2:\cdots : i_m):= \tau_{i_1:i_2} \tau_{i_2:i_3}\cdots \tau_{i_{m-1}:i_m} \tau_{i_{m}:i_1}.
\end{equation}
Nevertheless, this term is not a well defined $PSL(2,\mathbb{C})$ chain factors because there are extra powers of $\s_{n+1}$ and $\s_{n+2}$.  In order to fix this, we need to multiply the above expression by 
\begin{equation}
(n+1:n+2)^{(n-2)}=\left[ {-1\over 2^2\,y^{\rm T}_{n+1}\,y^{\rm T}_{n+2}}\left({1 \over y^{\rm T}_{n+1}-y^{\rm T}_{n+2}}\right)^2 \right]^{(n-2)}=
\left({1\over 2^4}\right)^{(n-2)}={1\over 2^{4n-8}}
\end{equation}

Putting \eqref{tauprod} and \eqref{chainfactor} together, the one-loop integrand takes the form
\begin{equation}\label{chainfactorSL}
\prod_{a}^n{1\over (\s_a\,y^{\rm T}_a)^2}=2^{8}\,\, {\prod_{a=1}^n(a:n+1)(a:n+2) \over (n+1:n+2)^{(n-2)}}  \, ,
\end{equation}
which is a well defined $PSL(2,\mathbb{C})$ integrand. 

To end, we show that the $(\s_n y_n^{\rm T})$ term is the tree level  Faddeev popov determinant, $\Delta_{\rm FP}(n,n+1,n+2)$.  Let us remember that the $PSL(2,\mathbb{C})$ generators over a quadratic curve were written in \cite{Gomez:2016bmv}. So, on the quadratic curve, $(y_a^{\rm T})^2=\s_a-1$,  these generators take the form 
\begin{equation}
L_1=2\sum_{a=1}^n y^{\rm T}_a \,\p_{\s_a}\, ,~~~ L_0=2\sum_{a=1}^n (y^{\rm T}_a)^2 \,\p_{\s_a}\, ,~~~  L_{-1}=2\sum_{a=1}^n (y^{\rm T}_a)^3 \,\p_{\s_a}\,\, ,
\end{equation}
with
\begin{equation}
[L_{\pm 1},L_0]=\pm \, L_{\pm 1}, ~~~ [L_{1},L_{-1}]=2\, L_0\, .
\end{equation}
Since the three fixed punctures are, $\s_n=c$, $\s_{n+1}=\s_{n+2}=0$, where $\s_{n+1}$ and $\s_{n+2}$  are located on different sheets, $y_{n+1}^{\rm T}=\sqrt{-1}={\rm I}$ and $y_{n+2}^{\rm T}=-\sqrt{-1}=-{\rm I}$, then the Faddeev-Popov determinant is given by
\begin{equation}
\Delta_{\rm FP}(n,n+1,n+2)=2^3
\left |
\begin{matrix}
y^{\rm T}_n & y^{\rm T}_{n+1} & y^{\rm T}_{n+2}\\
(y^{\rm T}_n)^2 & (y^{\rm T}_{n+1})^2 & (y^{\rm T}_{n+2})^2\\
(y^{\rm T}_n)^3 & (y^{\rm T}_{n+1})^3 & (y^{\rm T}_{n+2})^3\\
\end{matrix}
\right|
=2^3
\left |
\begin{matrix}
y^{\rm T}_n & ~\,\,{\rm I} & -{\rm I}\\
(y^{\rm T}_n)^2 & -1 & -1\\
(y^{\rm T}_n)^3 & -{\rm I} & ~\, \,{\rm I}\\
\end{matrix}
\right|
=
-2^4 \, {\rm I} \,(\s_n\,y^{\rm T}_n)\, ,
\end{equation}
therefore
\begin{equation}
2^8\,(\s_n\,y^{\rm T}_n)^2 = -\Delta^2_{\rm FP}(n,n+1,n+2).
\end{equation}

Now, we are ready to write the ${\rm A}_{n}^{\rm n-gon}$ integral as a tree level $\L$ prescription 
\begin{equation}\label{HyperSmatrixForwardtau}
{\rm A}_n^{\rm n-gon} =\int    {d^D\ell  \over\ell^2} \int_{\gamma} \left( \prod_{a=1}^n
{ y^{\rm T}_a\,d y^{\rm T}_a \over C^{\rm T}_a}\right)
\left(
\prod_{i=1}^{n-1}\,{d\s_i \over E^{ \rm T}_i}\right)
 {\Delta^2_{\rm FP}(n,n+1,n+2)\,\prod_{a=1}^n(a:n+1)(a:n+2) \over (n+1:n+2)^{(n-2)}},
\end{equation}
where we have drop the overall factor, ${\rm I}^{D}$, which is just a sign and it  does not affect the computation.

Note that if $\s_{n+1}=\s_{n+2}=0$ are on the same sheet, i.e.  $y^{\rm T}_{n+1}= y^{\rm T}_{n+2}=\sqrt{-1}={\rm I}$ (upper sheet) or  $y^{\rm T}_{n+1}= y^{\rm T}_{n+2}=-\sqrt{-1}=-{\rm I}$ (lower sheet), the factor
\begin{equation*}
\frac{\Delta^2_{\rm FP}(n,n+1,n+2)}{(n+1:n+2)^{(n-2)}}=2^6[-2^2\, y^{\rm T}_{n+1}\,y^{\rm T}_{n+2}(y^{\rm T}_{n+1}-y^{\rm T}_{n+2})^2]^{(n-2)}
\left |
\begin{matrix}
y^{\rm T}_n & y^{\rm T}_{n+1} & y^{\rm T}_{n+2}\\
(y^{\rm T}_n)^2 & (y^{\rm T}_{n+1})^2 & (y^{\rm T}_{n+2})^2\\
(y^{\rm T}_n)^3 & (y^{\rm T}_{n+1})^3 & (y^{\rm T}_{n+2})^3\\
\end{matrix}
\right|^2\, ,
\end{equation*}
vanishes trivially. This fact will imply  that these kind of configurations in the $\L$ algorithm will be zero, it will be shown in the next section \ref{Sec7}.  Finally, ${\rm A}_n^{\rm n-gon} $ can be written in terms of the $\L$ prescription of $(n+2)$ particles
\begin{equation}\label{HyperSmatrixForwardtau3}
{\rm A}_n^{\rm n-gon} =\int    {d^D\ell  \over\ell^2}\,\,  {\cal I}^{\rm n-gon}_n(1,2,\ldots , n |\ell,-\ell,),
\end{equation}
where we have defined the ${\cal I}^{\rm n-gon}_n(1,2,\ldots , n|\ell,-\ell)$ integrand as
\begin{align}\label{HyperSmatrixForwardtau2}
&{\cal I}_n^{\rm n-gon}(1,2,\ldots , n|\ell,-\ell) ={\cal I}^{\rm n-gon}_n(1,2,\ldots , n)\nonumber\\
& = \int_{\gamma} \left( \prod_{A=1}^{n+2}
{ y^{\rm T}_A\,d y^{\rm T}_A \over C^{\rm T}_A}\right)
\left(
\prod_{i=1}^{n-1}\,{d\s_i \over E^{ \rm T}_i}\right)
 {\Delta^2_{\rm FP}(n,n+1,n+2)\,\prod_{a=1}^n(a:n+1)(a:n+2) \over (n+1:n+2)^{(n-2)}},
\end{align}

Note that in the parcular case when the off-shell momenta, $k_{n+1}^\mu\neq -k_{n+1}^\mu $, but the momentum conservation is still satisfied,  $\sum_{A=1}^{n+2} k_A=0$, the integral
\begin{align}\label{HyperSmatrixForwardtauG}
{\cal I}_n(1,2,\ldots , n|i,j) = \int_{\gamma} \left( \prod_{A=1}^{n+2}
{ y^{\rm T}_A\,d y^{\rm T}_A \over C^{\rm T}_A}\right)
\left(
\prod_{a=1}^{n-1}\,{d\s_a \over E^{ \rm T}_a}\right)
 {\Delta^2_{\rm FP}(n, i, j)\,\prod_{b=1}^n(b:i)(b:j) \over (i:j)^{(n-2)}},
\end{align}
where we have denoted $\s_{n+1}:=\s_i$, $\s_{n+2}:=\s_j$, $k_{n+1}^\mu:=k_{i}^\mu$, $k_{n+2}^\mu:=k_{j}^\mu$ and the $ E^{\rm T}_a$'s scattering equations are given by 
\begin{equation}
 E^{\rm T}_a=
 \half\sum_{b=1\atop b\ne a}^{n} \left({y^{\rm T}_b\over y^{\rm T}_a}+1 \right) {k_b\cdot k_a \over \s_{ba}}   + \half \left({y^{\rm T}_i\over y^{\rm T}_a}+1 \right) {k_i\cdot k_a \over \s_{ia}}+ \half \left({y^{\rm T}_j\over y^{\rm T}_a}+1 \right) {k_j\cdot k_a \over \s_{ja}}
   \,, ~~ {\rm with}~~ a=1,\ldots n-1\,.\nonumber
\end{equation}
The above equations are not independent of the  gauge, i.e. if one fixes another puncture, for example $\s_{n-1}$, the final answer would be different. Nevertheles, when $k_{i}^\mu =-k_j^\mu=\ell^\mu$, then ${\cal I}_n(1,2,\ldots , n|\ell, -\ell):={\cal I}^{\rm n-gon}_n$ becomes gauge independent, as it was expected. We will see this explicitly on the final result.  We call ${\cal I}_n(1,2,\ldots , n|i,j)$ the  generalized ${\rm n-gon}$.

In addition, it has been shown in \cite{Gomez:2016bmv} that the integral in \eqref{HyperSmatrixForwardtau3}  can be rewritten in the CHY approach using the \eqref{mapping} transformation, $\s_a=z_a^2+1$,  as
\begin{equation}\label{CHYoneLoopn}
{\rm A}_n^{\rm n-gon} =\int {d^D\ell \over \ell^2}  \int\prod_{a=1}^{n-1}\,{dz_a\over E_a}{ 
(z_{n,n+1}z_{n+1,n+2}z_{n+2,n})^2\,
z_{n+1,n+2}^{2(n-2)}\over\prod_{b=1}^n
z_{n+1,b}^2z_{n+2,b}^2}\,,
\end{equation}
where $E_a,\,a=1,\dots,n-1$ are the scattering equations in the CHY approach. The inherited gauge fixing is given by $z_n=\sqrt{\s_n-1}, z_{n+1}={\rm I}, z_{n+2}=-{\rm I}$ and the forward limit $k^\mu_{n+1}=-k^\mu_{n+2}=-\ell^\mu,\,\ell^2\neq0$. 

It is worth to notice at this point that the expression above  coincides with the ${\rm n-gon}$ in \cite{Geyer:2015jch} and we think of it as a non trivial check of our results.

\section{$\Lambda$-algorithm}\label{Sec7}

In this section, we would like to use the $\Lambda$-algorithm developed in \cite{Gomez:2016bmv} in order to compute the integral \eqref{HyperSmatrixForwardtauG}, where we apply it directly on the generalized  ${\rm n-gon}$.  We will see that in the one loop case, the $\Lambda$-algorithm provide an expansion similar to  the Q-cuts found in \cite{Baadsgaard:2015twa}.  In this section we denote $(n+1)=i$ $(k_{n+1}=k_i)$ and $(n+2)=j$  $(k_{n+2}=k_j)$ for graphical convenience, be careful not to be confused with arbitrary particle indexes.

\subsection{Reviewing $\Lambda$-algorithm}

Let start by making a quick review of the $\Lambda$-algorithm by appliying it to \eqref{HyperSmatrixForwardtau2}. As we have seen in the section above, the integration over the moduli parameter $\lambda$ has left us with two Riemman spheres connected through a nodal fixed point such as the puntures get distributed among this two spheres. Roughly speaking, the almorithm implement a graphical way to sum over all possible distribution configurations of puntures allowed by the $PSL(2,\mathbb{C})$ symmetry on each individual sphere. For more details please refer to  \cite{Gomez:2016bmv}.

Before beginning it is useful to introduce the following notation
\begin{align}
&k_{a_1\ldots a_m}:=\sum_{a_i<a_j}^m k_{a_i}\cdot k_{a_j},\\
& [a_1,a_2,\ldots, a_m]=k_{a_1}+k_{a_2}+\cdots + k_{a_m},
\end{align}
and 
\begin{center}
\includegraphics[scale=0.5]{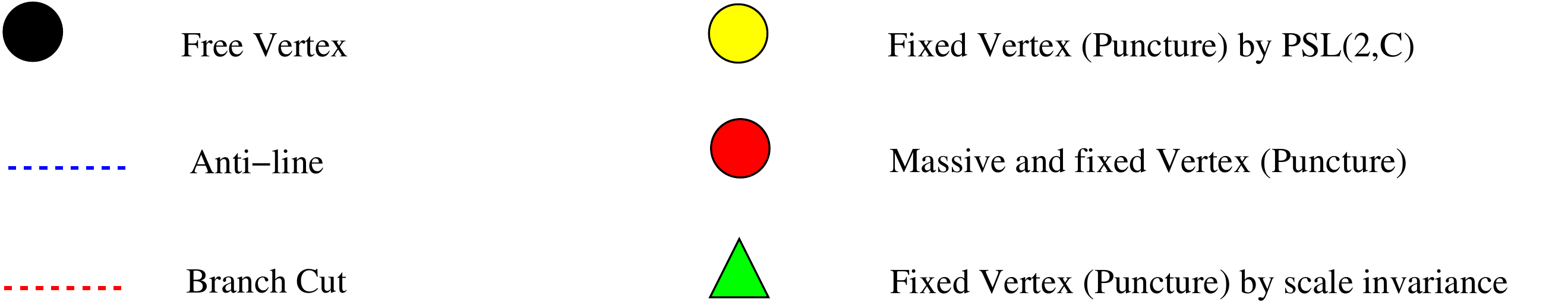}.
\begin{center}
Figure 2.:\,\,{\small {\rm Colored Vertices. \,}}
\end{center}
\end{center}

{\bf Steps}

\begin{itemize}
\item {\bf (1)}  Draw the graph corresponding to the integrand to be computed:\\
Each factor $\tau_{a:b}$ in the integrand is represented by a black solid line connecting vertex $a$ to vertex $b$, while each  factor $(\tau_{ab})^{-1}$ will be represented by a dotted blue line ({\bf anti-lines}).  Each vertex of the graph corresponds to a puncture. By $PSL(2,\mathbb{C})$ invariance, the number of solid black lines minus dotted blue lines coming into a given vertex should be equal to four. A given graph must have three yellow or red vertices representing the  fixed by $PSL(2,\mathbb{C})$ gauge freedom plus one green vertex fixed from the scale symmetry inherited from the scale invariance on the $\mathbb{CP}^2$ embedding
\begin{equation}\label{1loopintegrand}
\begin{aligned}
{\cal I}_{n}(123...n|i,j)~~=
\end{aligned}
\qquad
\raisebox{-35mm}{\includegraphics[keepaspectratio = true, scale = 0.3] {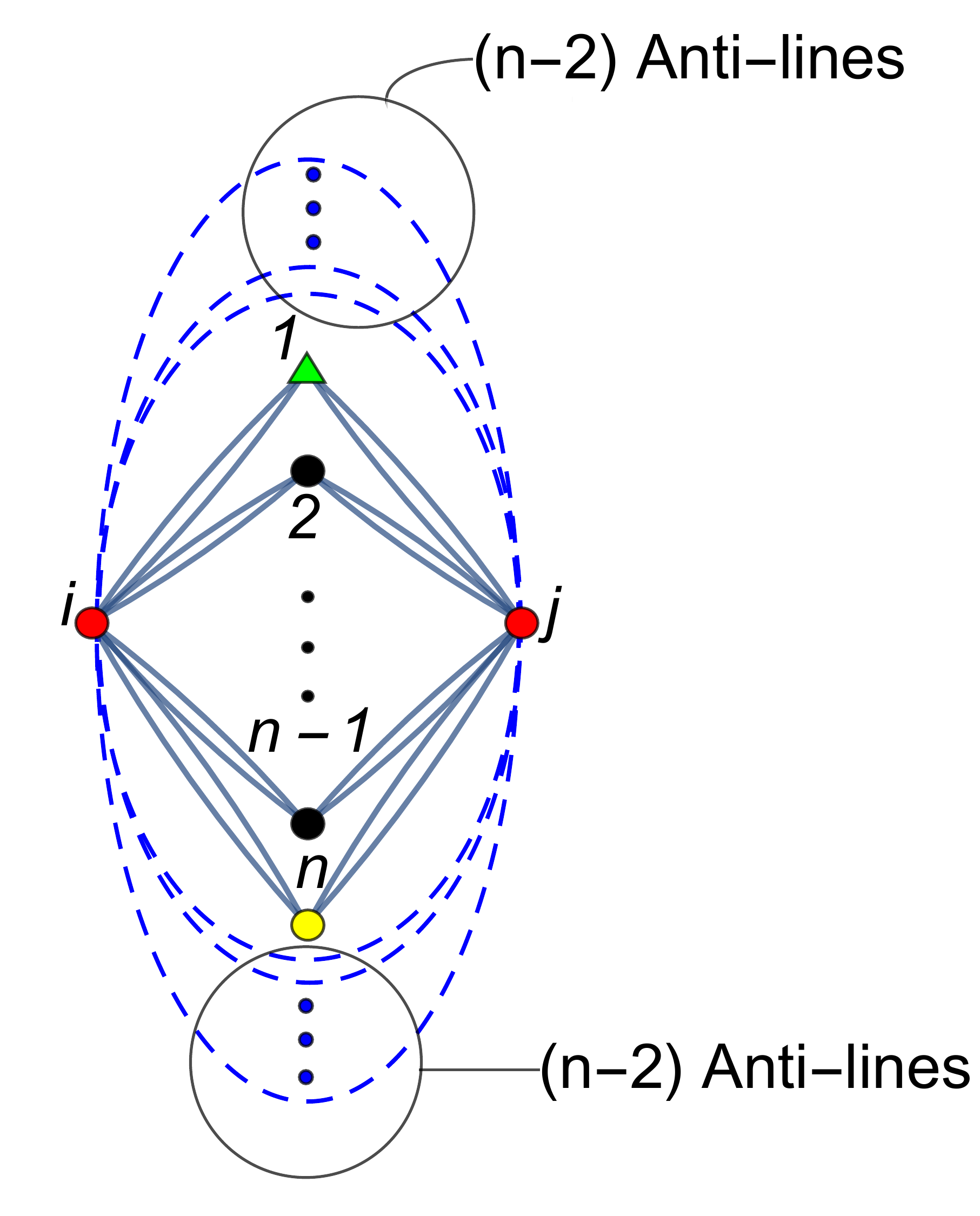}}
\end{equation}
\begin{center}
Figure 3:\,\,{\small {\rm  Graph of the Generalized {\bf n-gon}. \,}}
\end{center}

Notice that we are applying the $\L-$algorithm over the generalized ${\rm n-gon}$, ${\cal I}_n(123\ldots n|i,j)$. When, $k_i^\mu=-k_j^\mu=\ell^\mu$, then we obtain the original ${\rm n-gon}$ in \eqref{HyperSmatrixForwardtau2}
\begin{equation}
{\cal I}_n(123\ldots n|\ell, -\ell)={\cal I}^{\rm n-gon}_n(123\ldots n).
\end{equation}

\item {\bf (2)} Find all non-zero allowable configurations.  As we mention in the section above, after integration over the moduli, we end up with two sheets connected through a fixed point on each sheet. This means essentially that in order to have a $PSL(2,\mathbb{C})$ invariant configuration, we need to left two more fixed points on the ``upper" sheet as well as other two in the ``lower"one, as schematically represented in the following graph. Those will be called, allowable configurations. 
\begin{center}
\includegraphics[scale=0.22]{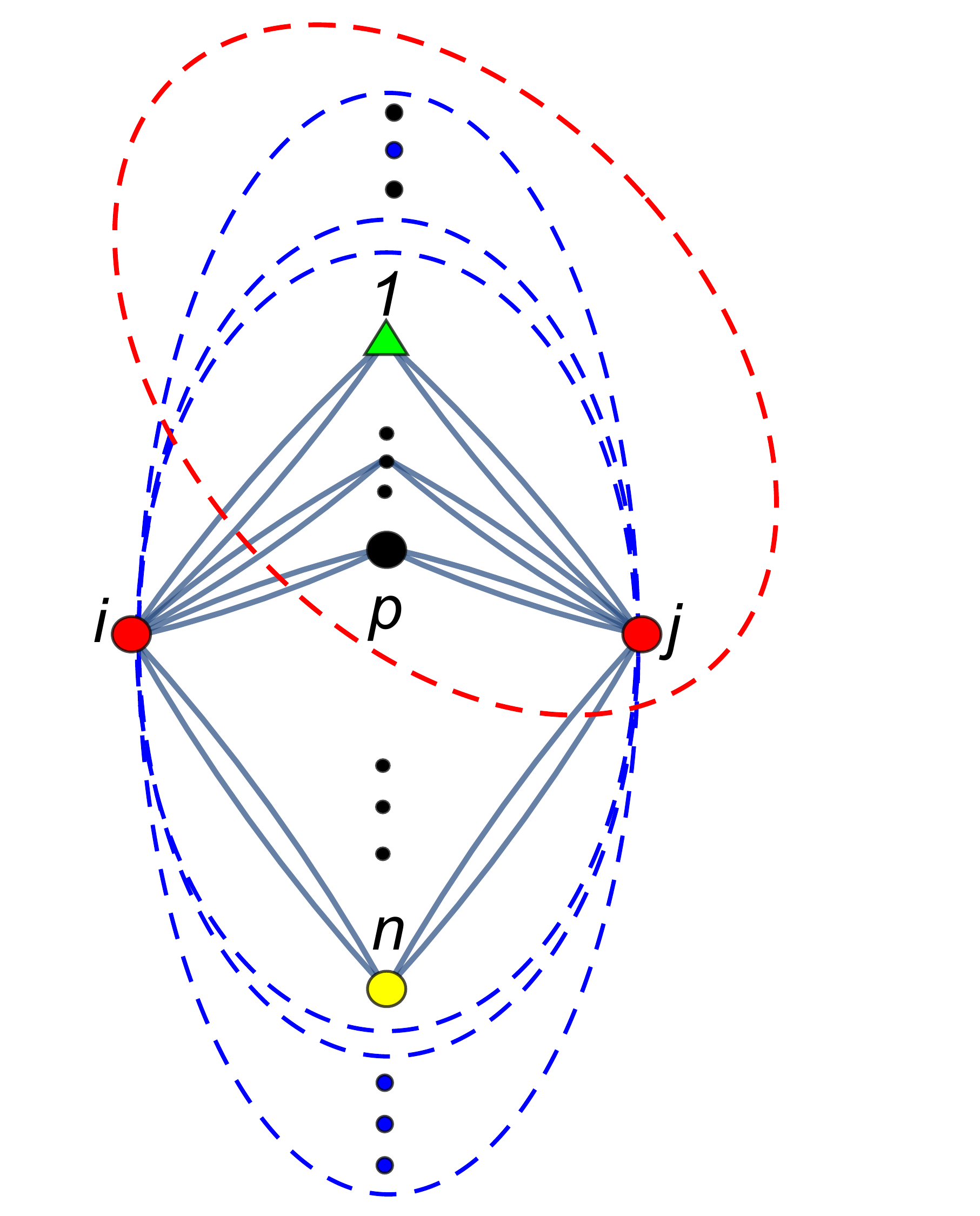}
\end{center}\begin{center}
Figure 4:\,\,{\small {\rm Example of a cut containing $p-$punctures. \,}}
\end{center}

Note that for this particular graph, the whole non-zero allowable configurations are also nonsingular configuration, i.e.
\begin{equation}
L-A=4,
\end{equation}
where $L$ is the number of lines and $A$ is the number of antilines which are intersected by the red line (branch cut).

\item {\bf (i)} The splitting is identified by a red line. The connecting point is then interpreted as two new (off-shell) punctures, one on the upper and the other one on the lower-sheet. The particles inside of a red line,  including now the new red massive puncture on the upper-sheet, shape a new graph on the upper-sheet  (subdiagram) and the particles  outside of the red line, including the new red massive puncture on the lower-sheet,  shape the another  graph  (subdiagram), such as it is shown in  figure 5.

The momentum of the red massive puncture on the upper-sheet is the sum over all momenta of the particles outside of the red line, i.e. 
\begin{equation}
k^{\rm upper}= [i,p+1,p+2,\ldots, n]
\end{equation}
and the momentum of the red massive puncture on the lower-sheet  is the sum over all momenta of the particles inside of the red line, i.e. 
\begin{equation}
k^{\rm lower}=[1,2,\ldots, p,j]  
\end{equation}

The scattering equation associates to the puncture in the green triangle, in our figure it is $E^{\rm T}_1$, becomes at the propagator
\begin{equation}\label{propa}
\frac{1}{E_1^{\rm T}}~\rightarrow ~ \frac{1}{k_{123\ldots p j}}.
\end{equation}
\begin{center}
\includegraphics[scale=0.27]{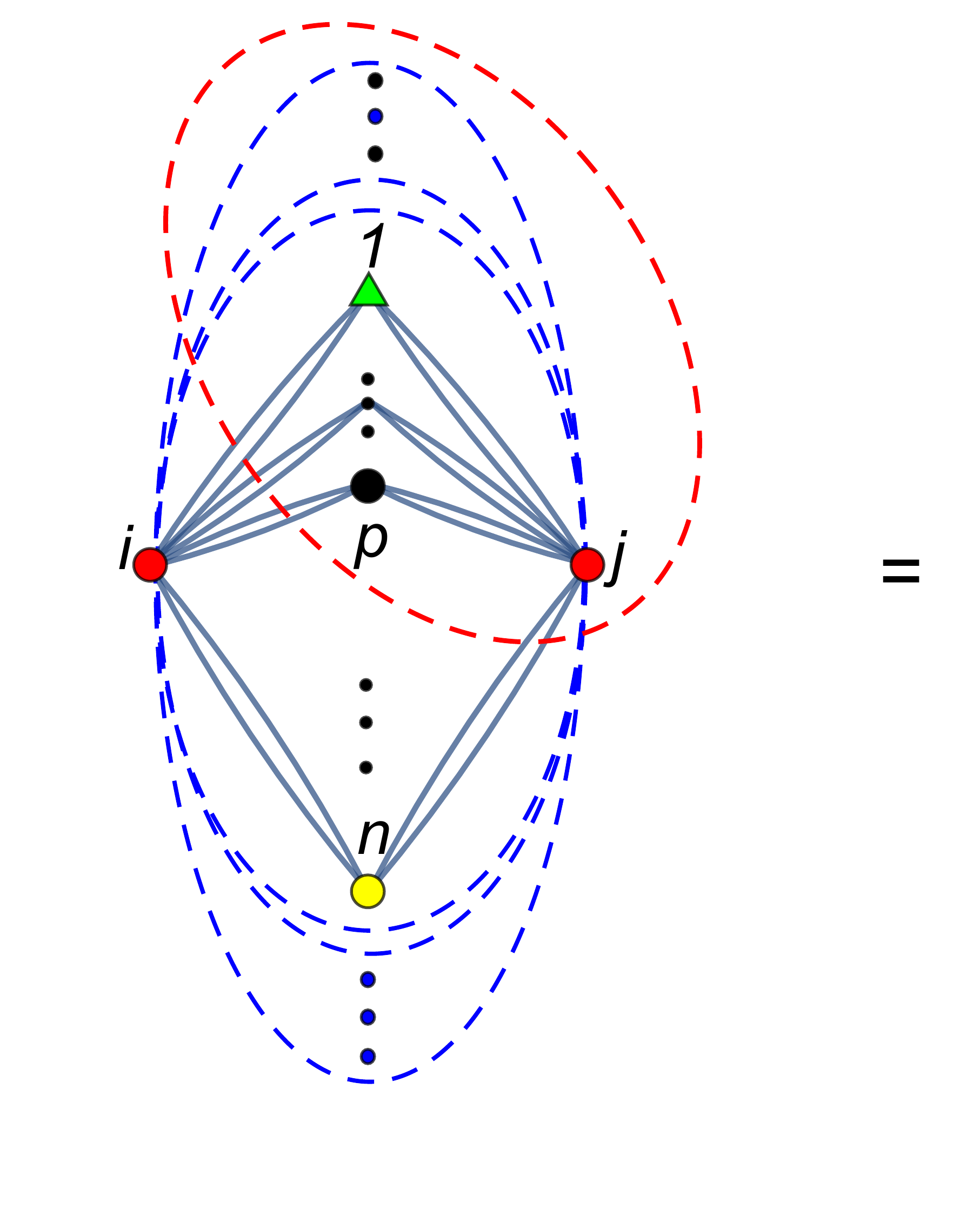}
\includegraphics[scale=0.25]{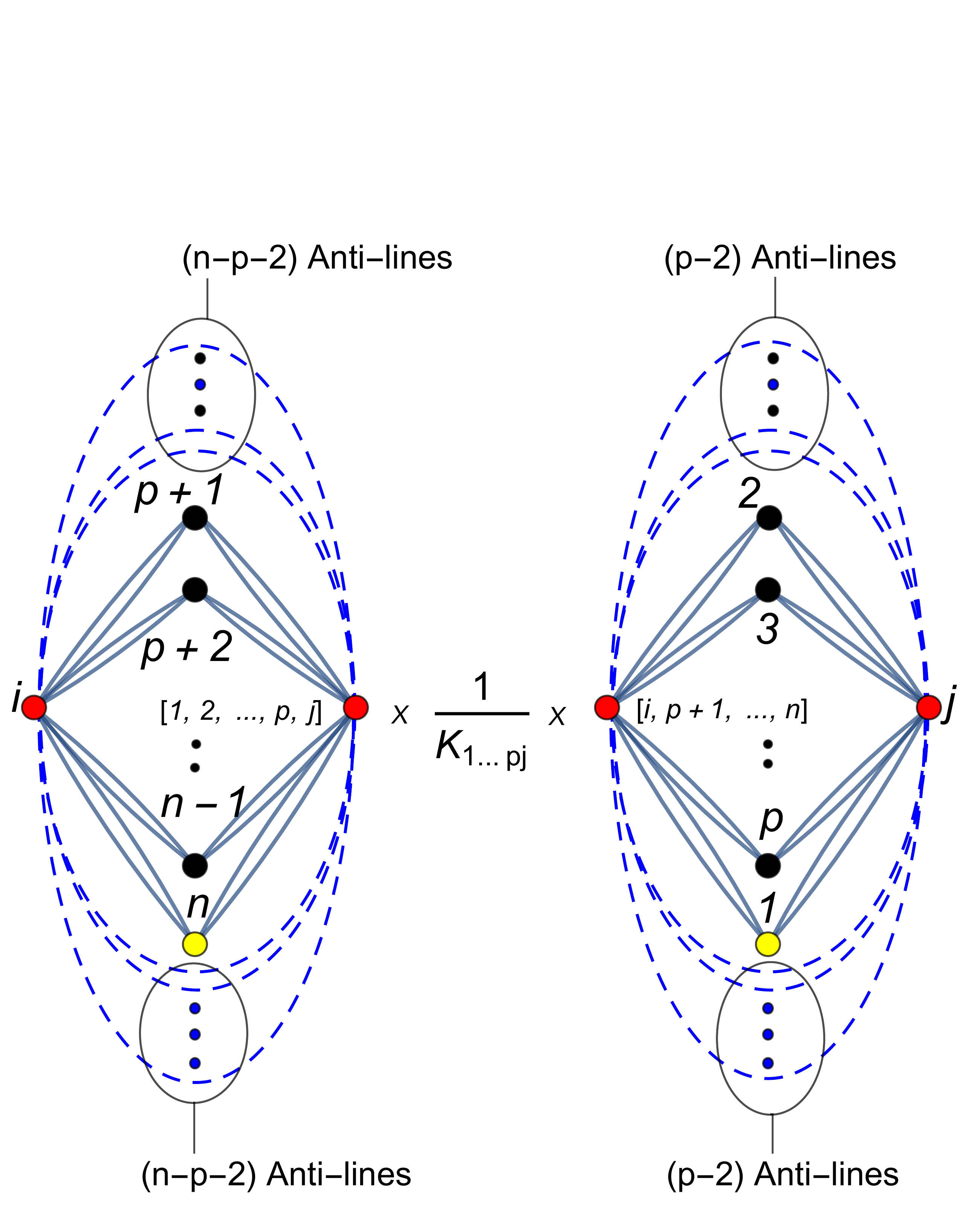}
\end{center}\begin{center}
Figure 5:\,\,{\small {\rm Example of a cut containing $p-$punctures and its decomposition. \,}}
\end{center}

Finally, note that the two new subdiagrams  are given in the original CHY approach, where $(\s_{i,p+1,\ldots n,},\s_j,\s_1)$ are the gauged punctures on the upper-sheet and  $(\s_{1,\ldots,p,j},\s_i,\s_n)$ are the gauged punctures on the lower-sheet.

\item {\bf (3)} Come back to the step (1). 

Keep performing the cutting of the sub diagrams until the lowest possible non-trivial diagram is reached.
It is useful to remember that  a 4-regular graph with 3 vertices is equals to one. 
\begin{center}
\includegraphics[scale=0.21]{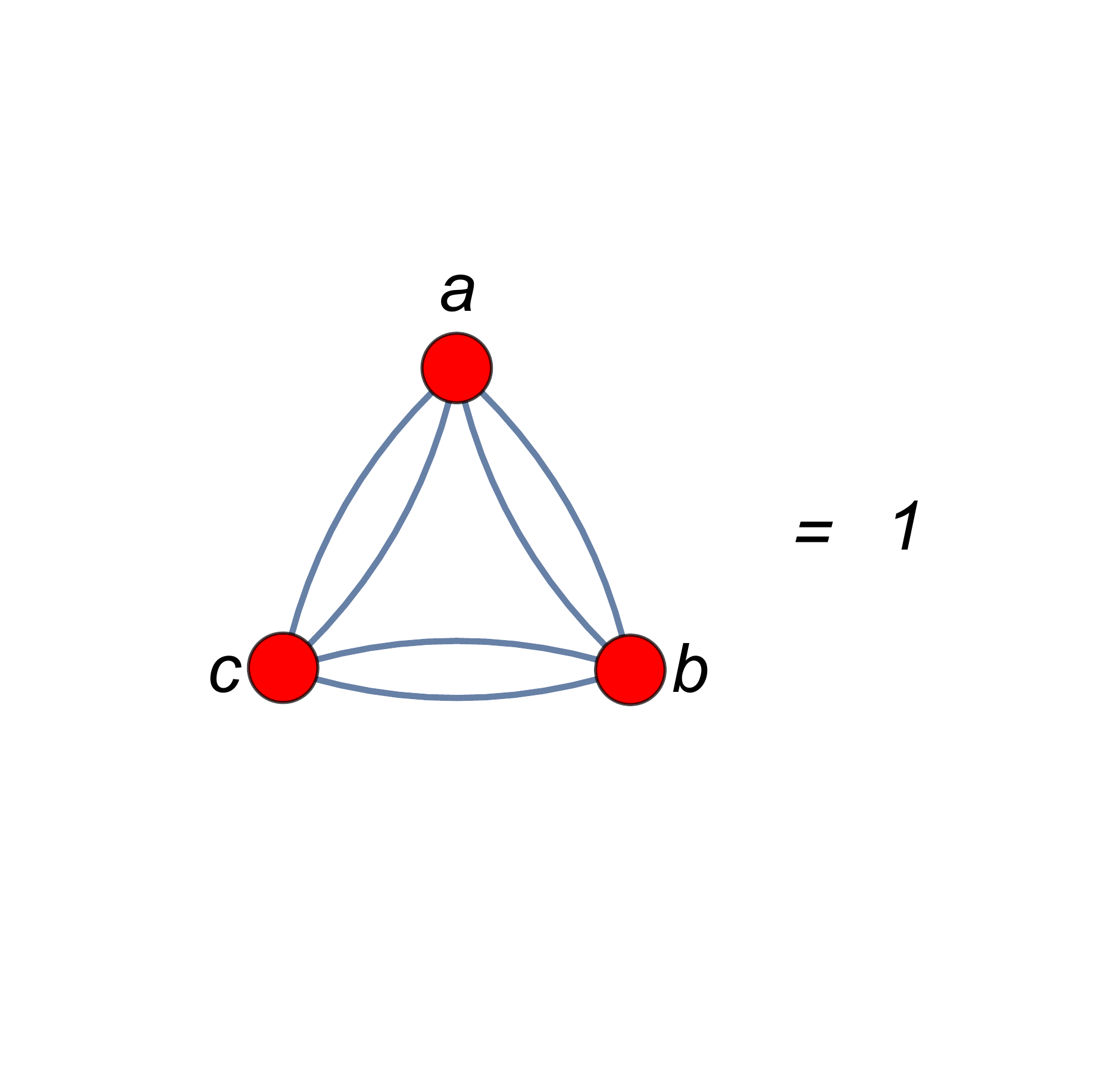}
\begin{center}
Figure 6:\,\,{\small {\rm 3-point 4-regular graph\,}}
\end{center}
\end{center}

\end{itemize}

{\bf Remark}

At this point let us recall that in order to formulate the CHY prescription at loop level Cachazo, He and Yuan \cite{Cachazo:2015aol}\footnote{Recently generalized to higher loops in \cite{Feng:2016nrf}} have considered the tree-level formalism with $n+2$ particles, such as the two extra particles are taken in the forward limit, namely $k^{\mu}_{n+1}\,(k^\mu_i)=-k^{\mu}_{n+2}\,(k^\mu_j)=\ell^{\mu}$. As we have seen in section \ref{factorization}, both the raising of the two-extra particles as well as the forward limit are encoded naturally in the elliptic scattering equations.

Notice also that if we take $i$ and $j$ ($\s_{n+1}$ and $\s_{n+2}$) at the same sheet in (\ref{treeSE}), the explicit dependence on $\ell$ in the scattering equations cancel out and they become those of $n-$particles at tree level. From the graphical point of view of the $\Lambda-$algorithm, we see that in such a case, i.e. when we cut the diagram in such a way that the punctures $i$ and $j$ end up over the same sheet, then the propagator (\ref{propa}) connecting the two sub-diagrams does not contains the off-shell momentum $\ell$ and the propagator becomes the usual factorization pole expected when a subset of punctures approach to a single point.  However, as we have shown in section \ref{Agon}, the Faddeev-Popov determinant vanish on those configurations and therefore they do not contribute to the $A^{{\rm n-gon}}$ integral. 
On the other hand, when the punctures corresponding to $k_i=-k_j=\ell$ are localized on different sheets, as is the case producing the loop result, the propagator connecting the sheets becomes
$$1\over{(\sum_{i=0}^p k_{a_i})^2+2\ell\cdot(\sum_{i=0}^pk_{a_i})}$$
which is the proper pole expected form the Q-cuts expansion\footnote{This does not mean that the expansion have to coincides with the Q-cut expansion.}.

The above discussion connects nicely with the analysis at section 4.1 in \cite{Geyer:2015jch} (see also \cite{Cachazo:2015aol}). There, by studying the resulting scattering equations on the factorization channels, the authors have reached the same conclusions we have realized in the above discussion. The given factorization is also naturally contained in our formalism.

\subsection{The ${\rm n-gon}$ and a new  recurrence relation}\label{ngonR}

Since the $\Lambda$-algorithm is graphical in nature, let us consider a few of the graphics decomposing configurations, figure 5, out from the total $2^{(n-1)}$, (see below for the counting of the total amount of diagrams). There are several types of different cuts. They can be first classified by the amount of unfixed punctures (black dots) inside the cut. The simplest example of one of such cuts with its decomposition is

\raisebox{1mm}{\qquad\qquad\includegraphics[scale=0.25]{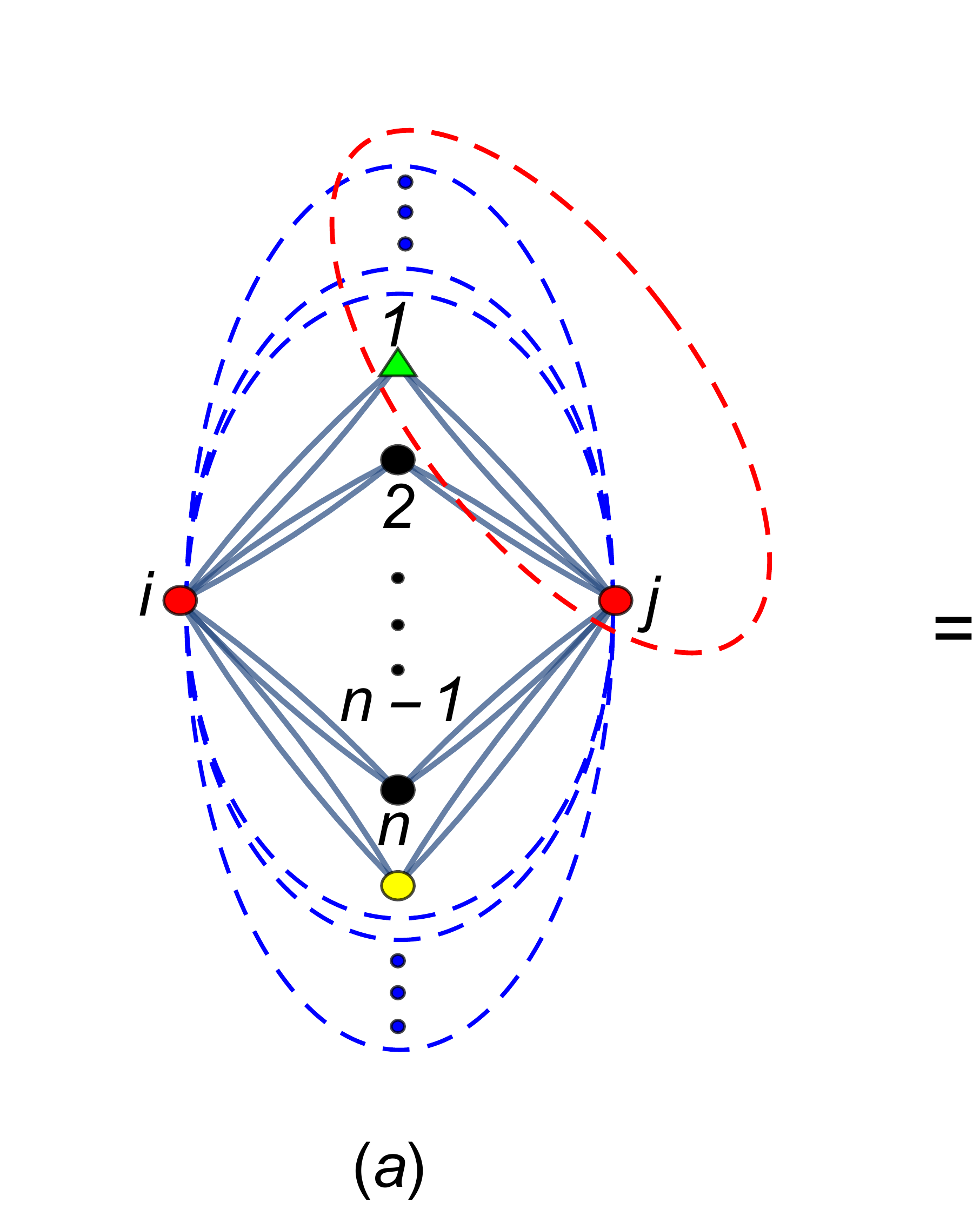}}
\raisebox{1mm}{\includegraphics[keepaspectratio = true, scale = 0.25] {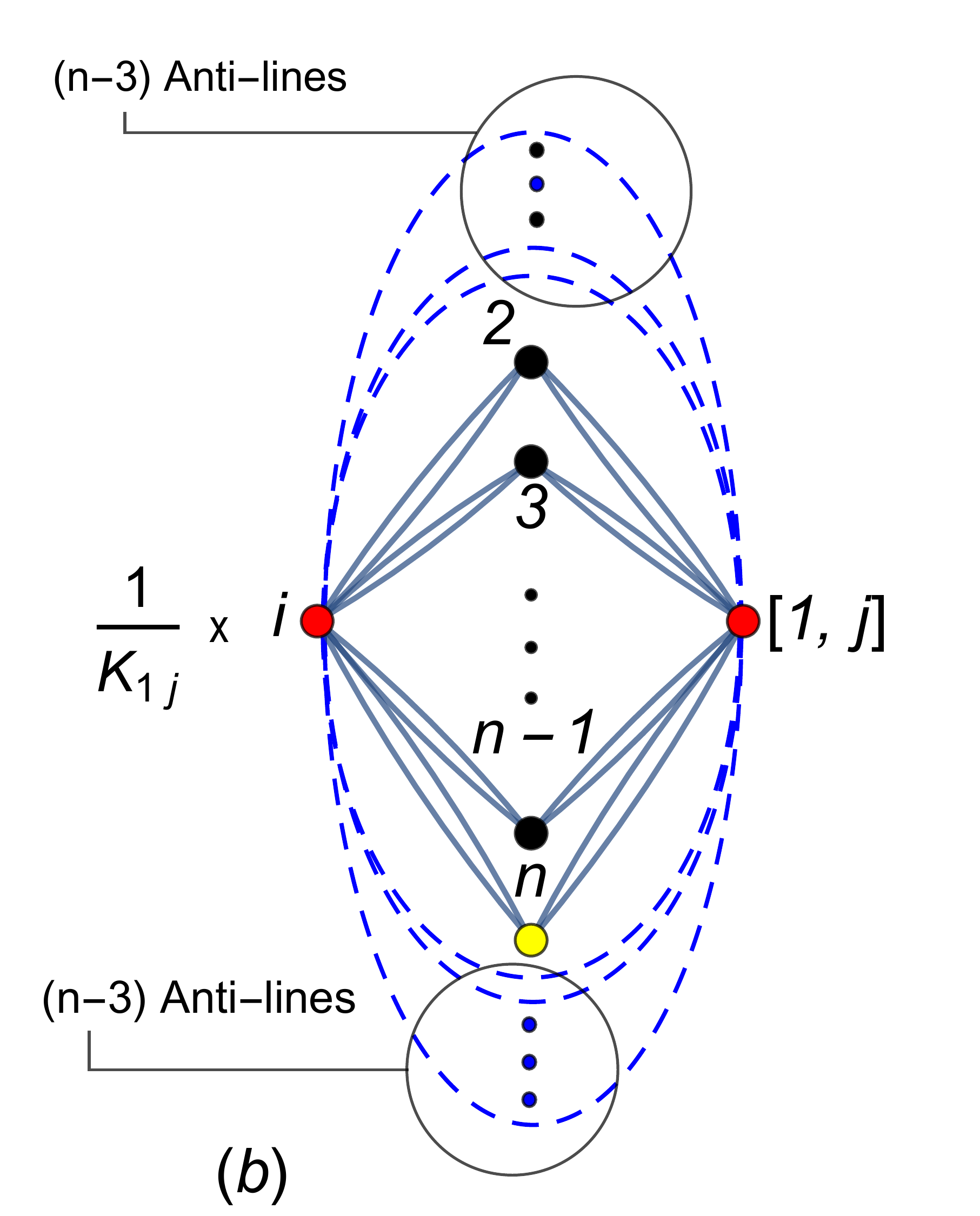}}.
\begin{center}
Figure 7:\,\,{\small {\rm {\bf (a)} Simplest allowed non-zero configuration. {\bf (b)} Result for the configuration in (a). \,}}
\end{center}
A more illustrative  examples of different kind of cuts containing more than two fixed punctures is presented in the following figure,
\begin{center}
\includegraphics[scale=0.23]{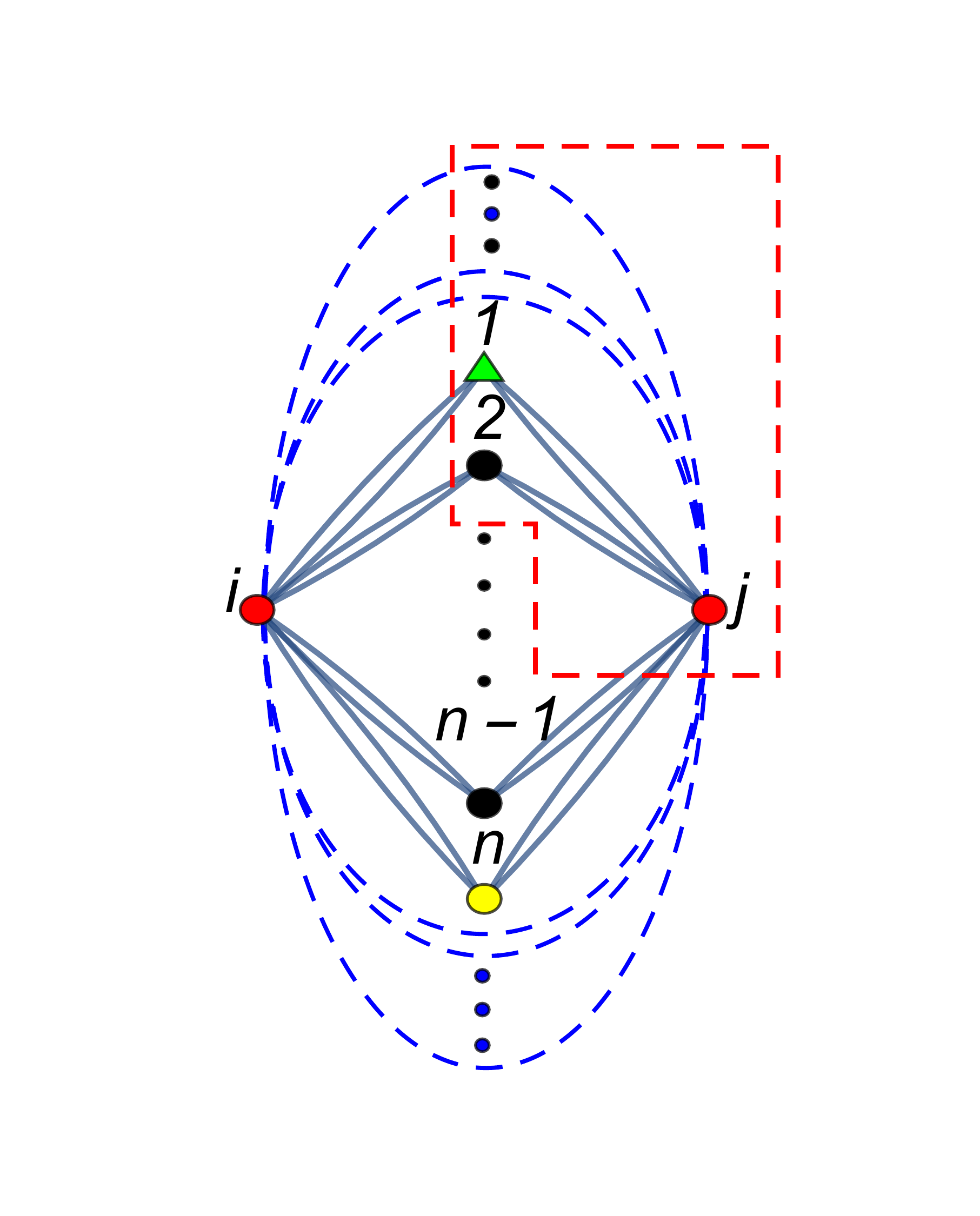}
\includegraphics[scale=0.23]{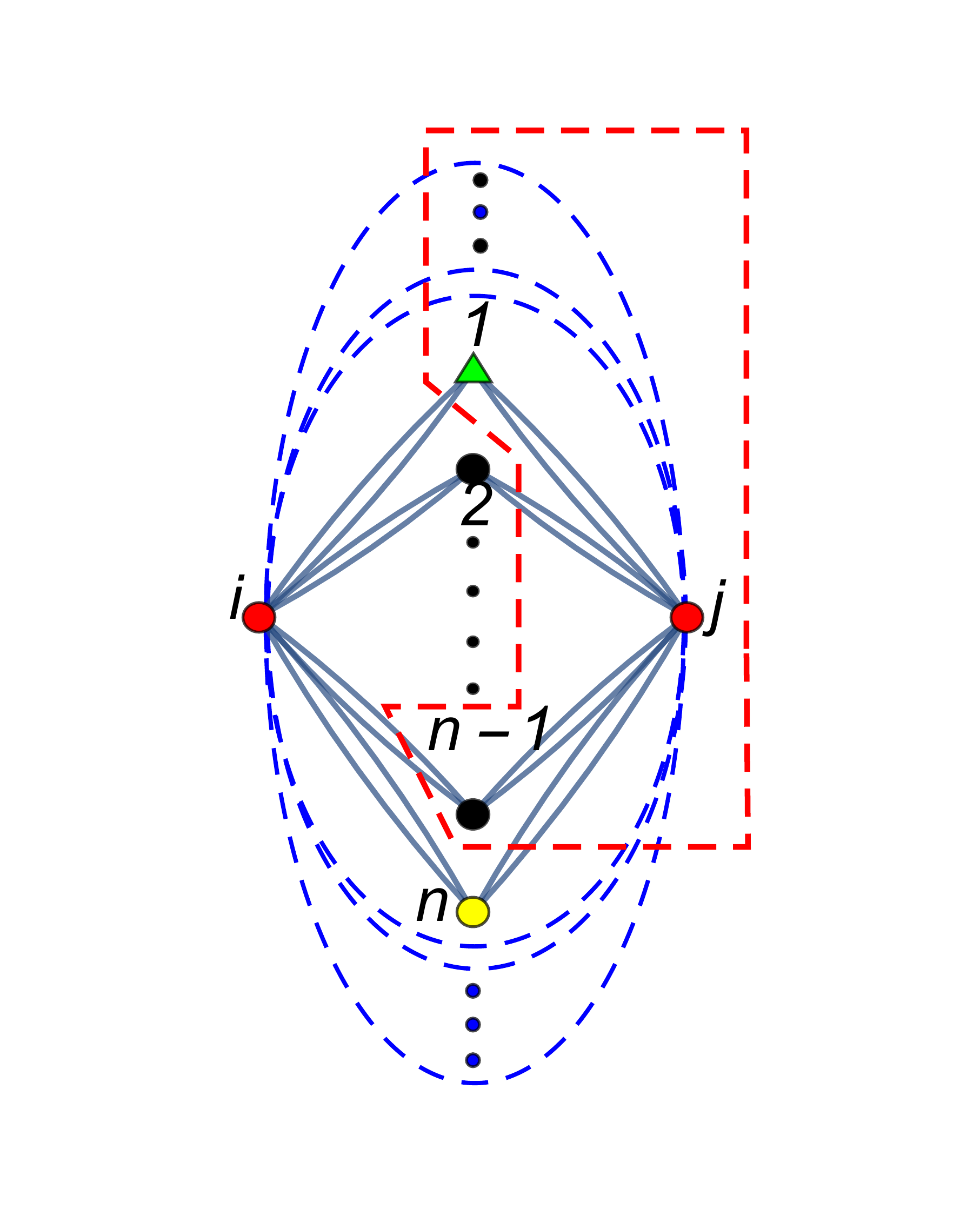}
\includegraphics[scale=0.23]{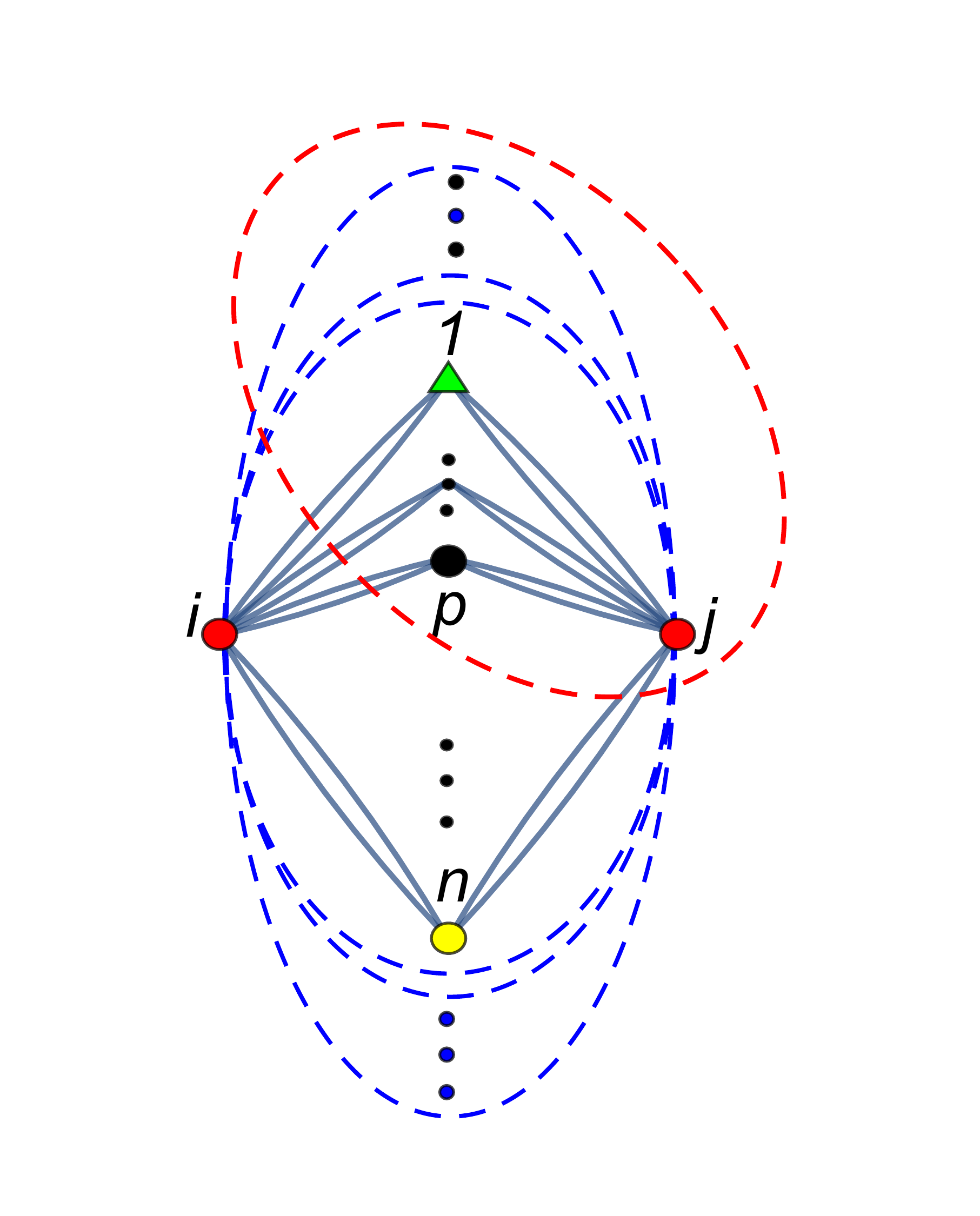}
\end{center}\begin{center}
Figure 8:\,\,{\small {\rm Example of cuts containing more than two fixed punctures. \,}}
\end{center}
The examples above are enough to deduce the remaining decompositions of the integrand (\ref{1loopintegrand}). Let $D_{p}$ be the number of  non-zero allowable configurations whose cut include $p-$unfixed punctures. For instance, note that $D_0=2$ since the configurations
\begin{center}
\includegraphics[scale=0.25]{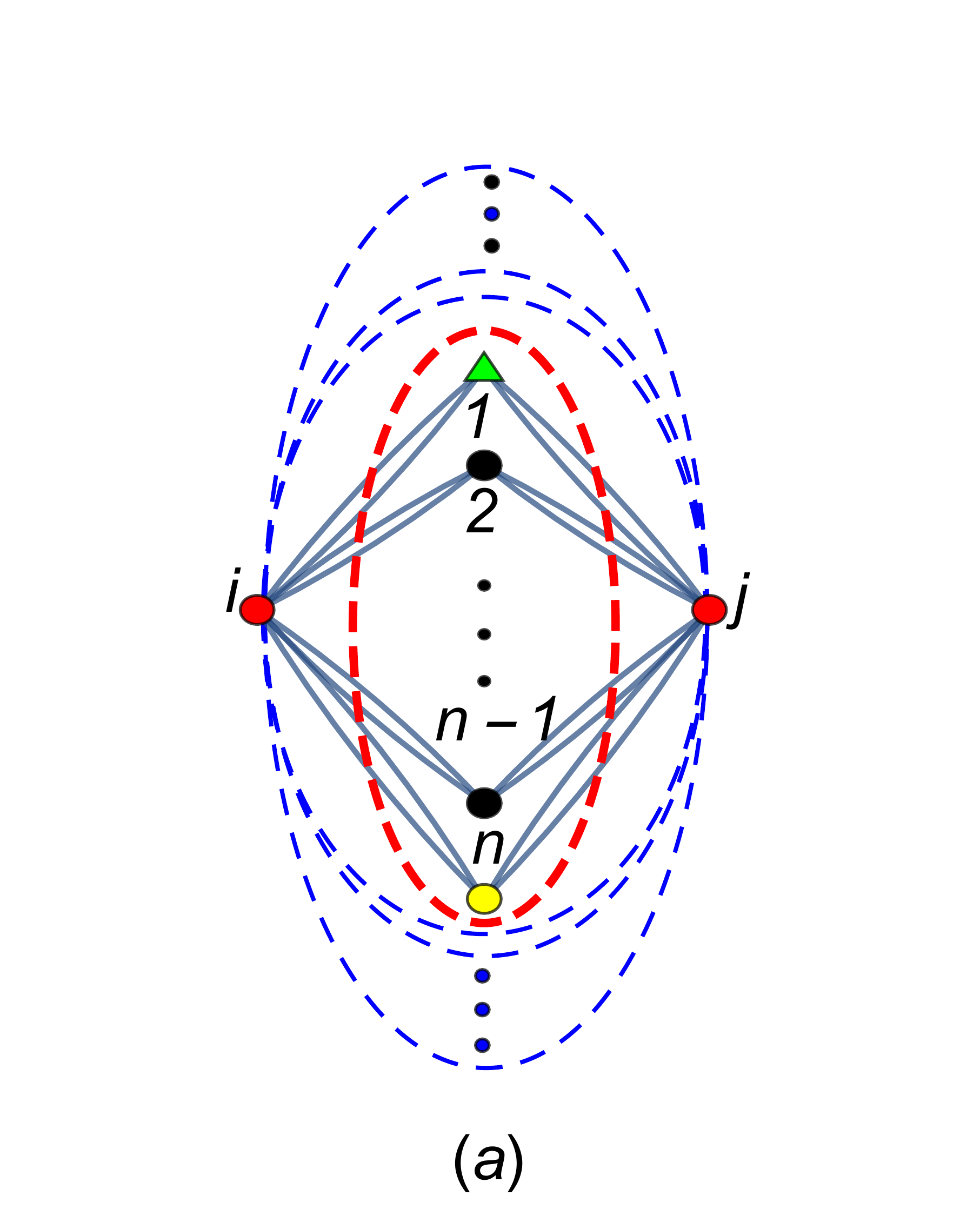}
\includegraphics[scale=0.25]{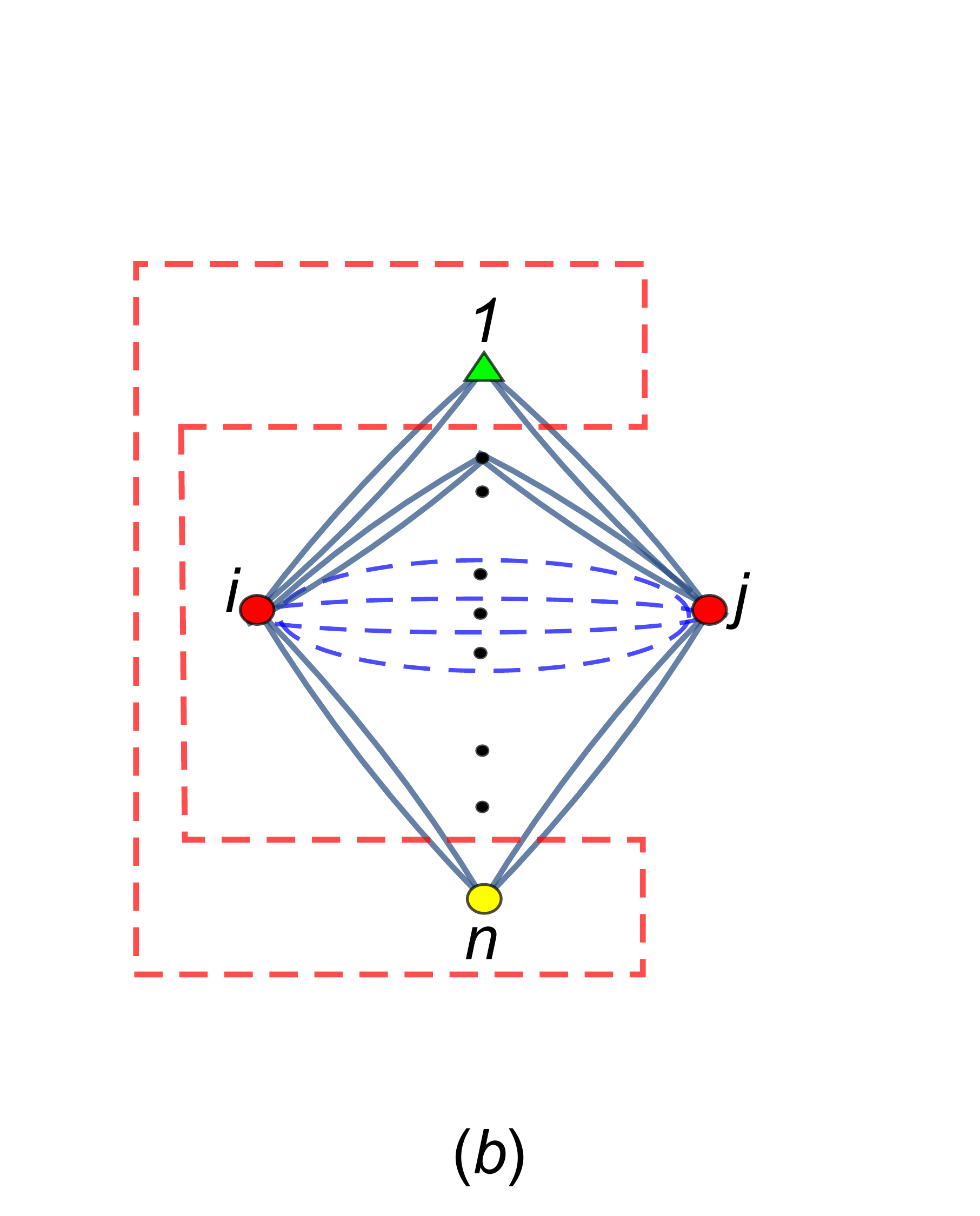}
\end{center}\begin{center}
Figure 9:\,\,{\small {\rm Allowable configurations which vanish trivially. \,}}
\end{center}
vanish trivially by the $\L$ theorem given in \cite{Gomez:2016bmv}. Clearly the configuration in figure (9a) is the same cut  as one given by the red line only encloses the punctures (i) and  (j). In fact, the all posibles configurations where the punctures (i) and (j) are on the same sheet vanish trivially by the $\L$ theorem. So, it is straightforward to check  
\begin{equation}
D_p=
2\times
\begin{pmatrix}
n-2 \\
p
\end{pmatrix},
\end{equation}
where the combinatorial number
coming from the different ways to pick $p$ unfixed points out from $(n-2)$ and the number  two coming from the interchange of $j$ by $i$.
This allows us to compute the total number of non-zero allowable configurations (${\rm T.N.A.C.}$) corresponding to apply the $\Lambda-$algorithm over (\ref{1loopintegrand})
\begin{equation}\label{totaldecomp} 
{\rm T.N.A.C.}=\sum_{p=0}^{n-2}D_p=
2\sum_{p=0}^{n-2}
\begin{pmatrix}
n-2 \\
p
\end{pmatrix}
=2^{n-1}\, .
\end{equation}

As we see from Figure 5, each configuration given by a cut which encloses $(p-1)$-unfixed punctures  splits in two smaller graphs of the same form as (\ref{1loopintegrand}), one with $p+2$ punctures and the other with $n-p+2$, which lead us to a nice recurrence relation. 

Before writing  this new recurrence relation we give some definitions which will be useful. Let $s_p$ be the set of $p$ ordered elements, i.e.
\begin{equation}
s_p:=\{a_1,a_2,\dots , a_{p}\},~~ {\rm where}~~ a_1<a_2<\dots <a_{p}\,~~{\rm and }~~ a_i\in\{2,\ldots, n-1\}. 
\end{equation}
Note that $p=0,\ldots, n-2$ and $s_0=\emptyset$.  We also define $S(p)$ as the set of $s_p$ elements, that is
\begin{equation}
S(p):=\{{\rm All~ possibles~} s_p \},
\end{equation}
for example,
\begin{equation}
S(0)=\{\emptyset\}, \qquad  S(1)=\{\{2\},\{3\}\ldots , \{ n-1 \}  \}.
\end{equation}
Finally, we denote  $\hat{s}_p$ as the ordered complement of $s_p$
\begin{equation}
\hat{s}_p:=\{b_1,\ldots ,b_{n-2-p}\},~ {\rm with}~ b_1<\dots <b_{n-2-p},~{\rm such~that }~ s_p\cup \hat{s}_p = \{ 2,\ldots, n-1 \},
\end{equation}
for instance
\begin{equation}
\hat s_0=\{2,\ldots , n-1  \}, \qquad \hat s_{n-2}=\emptyset.
\end{equation}

With these definitions, the recurrence relation expansion looks like,
\begin{align}\label{Expan1}
&{\cal I}_{n}(1,2,\dots ,n | i,j)=\\
&\sum_{p=0}^{n-2}\sum_{s_p\in S(p)}
{{\cal I}_{n-p-1}(\hat{s}_{p},n\,|\, i,[1,s_{p},j])\times{\cal I}_{p+1}(s_{p},1\,  |\, [i,\hat{s}_{p},n],j)\over k_{1, s_{p}, j}}+(i\leftrightarrow j)\, ,\nonumber
\end{align}
where  
$$
{\cal I}_1(a|i,j) =1,
$$
which is the 3-point function given in figure (6).

By taking, $i=-\ell$, $j=\ell$,  we obtain the  recurrence relation for the ${\rm n-gon}$ integrand
\begin{align}\label{Expangon}
&{\cal I}^{\rm n-gon}_{n}(1,2,\dots ,n) = \\
&\sum_{p=0}^{n-2}\sum_{s_p\in S(p)}
{{\cal I}_{n-p-1}(\hat{s}_{p},n\,|\, -\ell,[1,s_{p},\ell])\times{\cal I}_{p+1}(s_{p},1\,  |\, [-\ell,\hat{s}_{p},n],\ell)\over k_{1, s_{p}, \ell}}+(-\ell\leftrightarrow \ell)\, ,\nonumber
\end{align}
Although the above recurrence relation looks like the Q-cut expansion discussed in \cite{Huang:2015cwh, Baadsgaard:2015twa}, we conjecture that it is in fact the partial fraction expansion, i.e.
\begin{align}\label{pfe}
&\sum_{p=0}^{n-2}\sum_{s_p\in S(p)}
{{\cal I}_{n-p-1}(\hat{s}_{p},n\,|\, -\ell,[1,s_{p},\ell])\times{\cal I}_{p+1}(s_{p},1\,  |\, [-\ell,\hat{s}_{p},n],\ell)\over k_{1, s_{p}, \ell}}+(-\ell\leftrightarrow \ell)\, \nonumber\\ 
&=
\sum_{\s\in S_n}\frac{1}{\ell\cdot k_{\s1}(\ell\cdot (k_{\s1}+k_{\s2}) +k_{\s1}\cdot k_{\s2} )  \cdots (-\ell\cdot k_{\s n} ) },
\end{align}
on the support of momentum conservation constraint, $\sum_{a=1}^n k_a^\mu=0$, where  $S_n$ is the permutation group of $n$ elements.
This is not a trivial result and we do not have a proof of it. In addition, we have checked this conjecture numerically up to 9 points and in an analytical way up to 4 points.

It is worth remembering that the partial fraction expansion in \eqref{pfe} coming from the  Feynman diagram ${\rm n-gon}$, given by
\begin{equation}
\ell ^2\,\,{\cal I}^{\rm n-gon}_{\rm Feynman}= \frac{1}{(\ell+ k_{\s1})^2(\ell +k_{\s1}+k_{\s2})^2  \cdots (-\ell+ k_{\s n} )^2 }\, ,
\end{equation}
after using the partial fractions identity \cite{Feynman:1963},
\begin{equation}\label{partialfrac}
{1\over \prod_{i=1}^n D_i}=\sum_{i=1}^{n}{1\over D_i\prod_{j\neq i}(D_j-Di)}\,. 
\end{equation}

At this point we would like to stress that the $\Lambda-$algorithm can indeed do better.  For instance, despite the Q-cut expansion allow us to rewrite the ${\rm n-gon}$ as a sum over lower off-shell sub-amplitudes, namely ${\cal I}_{p},\,{\cal I}_{n-p}$, for \eqref{Expan1} and \eqref{Expangon} we can keep using the $\Lambda-$algorithm over those subamplitudes to rewritte each of then as even lower graphs until we reach the lowest sub-amplitude. Clearly,  the lowest sub-amplitude is  given by the generalized ${\rm 2-gon}$, ${\cal I}_{2}(a,b|i,j)$,

\begin{equation}\label{bubble}
\begin{aligned}
{\cal I}_{2}(a,b|i,j)~~=
\end{aligned}
\qquad
\raisebox{-30mm}{\includegraphics[keepaspectratio = true, scale = 0.25] {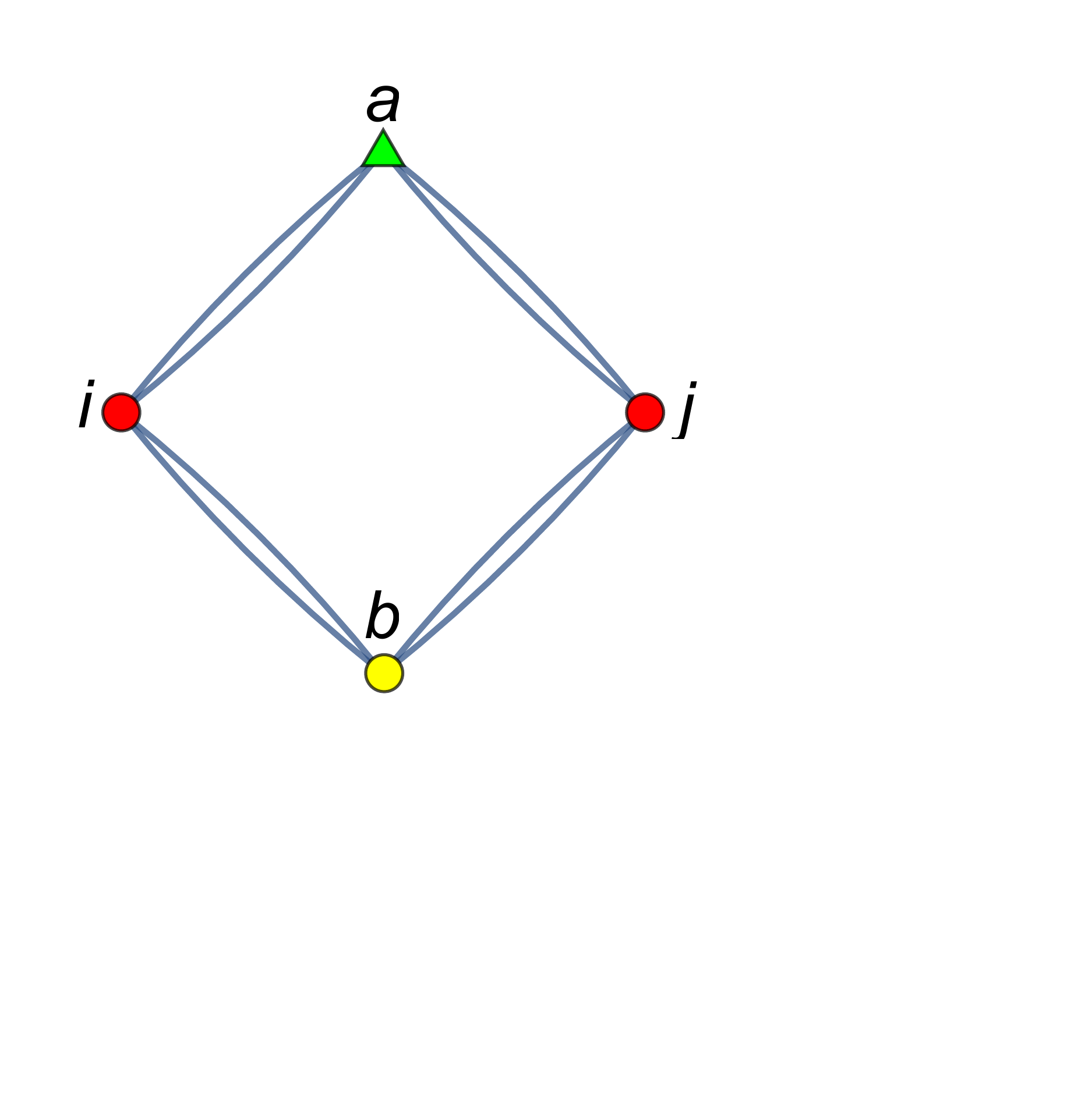}},
\end{equation}
\begin{center}
Figure 10:\,\,{\small {\rm Generalized {\bf 2-gon}. Fundamental building block. \,}}
\end{center}
which corresponds to the bubble Feynman diagram. Hence, the whole expansion (\ref{Expan1}) atomize to a sum of bubble diagrams only.
As we can see immediately from using the $\Lambda-$algorithm , the diagram in \eqref{bubble} is solved in a simple way
\begin{center}
\includegraphics[scale=0.28]{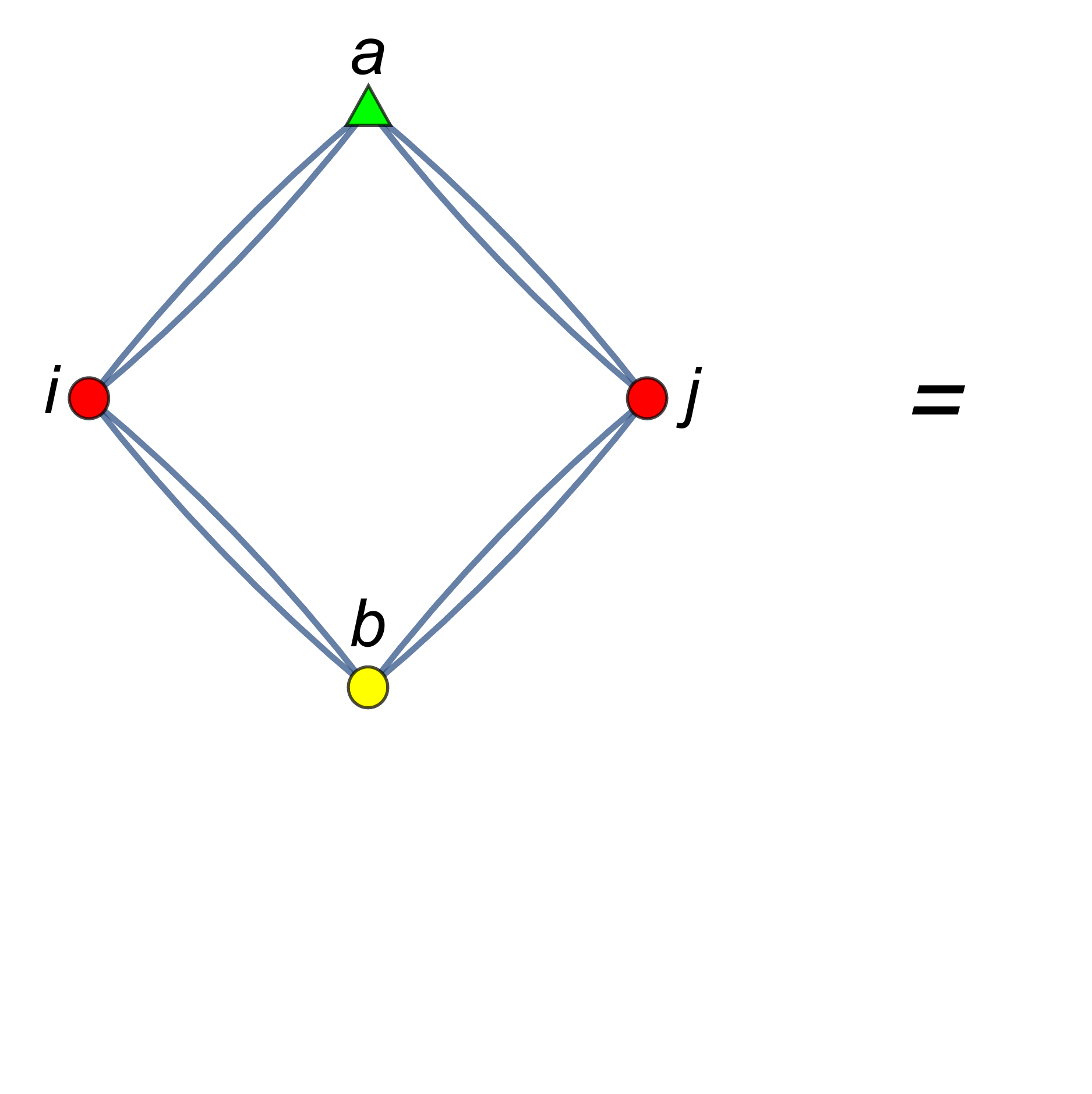}
\includegraphics[scale=0.28]{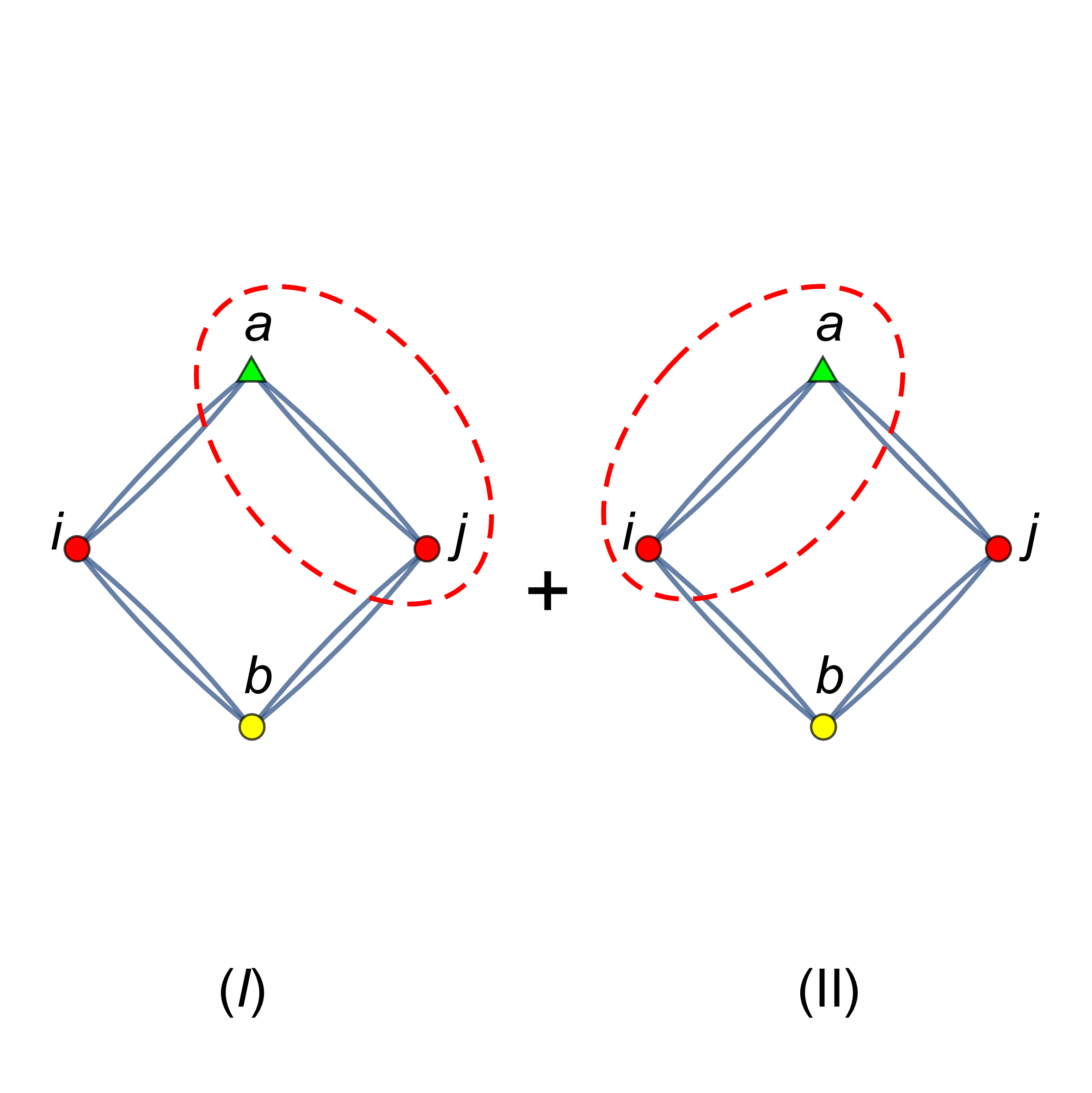}
\begin{center}
Figure 11: \,{\small {\rm All possible non-zero configurations for the  Bubble.\,}}
\end{center}
\end{center}
\begin{equation}\label{Gbubble}
 {\cal I}_2 (a,b|i,j)={1\over k_{aj}}+{1\over k_{ai}}\,.
 \end{equation}
Thus, Schematically the (\ref{Expan1}) expansion should have the following form,
\begin{align}\label{ultimate}
&{\cal I}_{n}^{\rm n-gon}(a_1,a_2,\dots, a_n|-\ell,\ell)=\nonumber\\
&\sum_{(s_{p_1},s_{p_2},s_{p_3})\in S(p)}\prod_{(s_{p_1},s_{p_2},s_{p_3})}
{{\cal I}_{2}(a_r,a_q|[s_{p_1},-\ell] , [s_{p_2},\ell])\over 
k_{s_{p_3}\ell}} + (\ell \leftrightarrow  - \ell)    \,. 
\end{align}
By (\ref{ultimate}), we means one can sum over the product of all the possible ${\cal I}_{2}$'s divided by all possible $k_{s_{p_3}\ell}$'s. Of course not all such a terms will contribute to the final result and there is many redundancies, but we want to use the above expression as a way to display just the form of the final answer. This expansion will be clarified through the simplest examples discussed in the following section. 

\section{Lower point examples}\label{Sec8}

For the sake of clarity we would like to show in this section some particular examples for the scattering of three and four scalar particles at one loop.
\subsection{Three-particles scattering} Following the discussion at the section above the one-loop $n=3$ graph is given by,
\begin{center}
\includegraphics[scale=0.2]{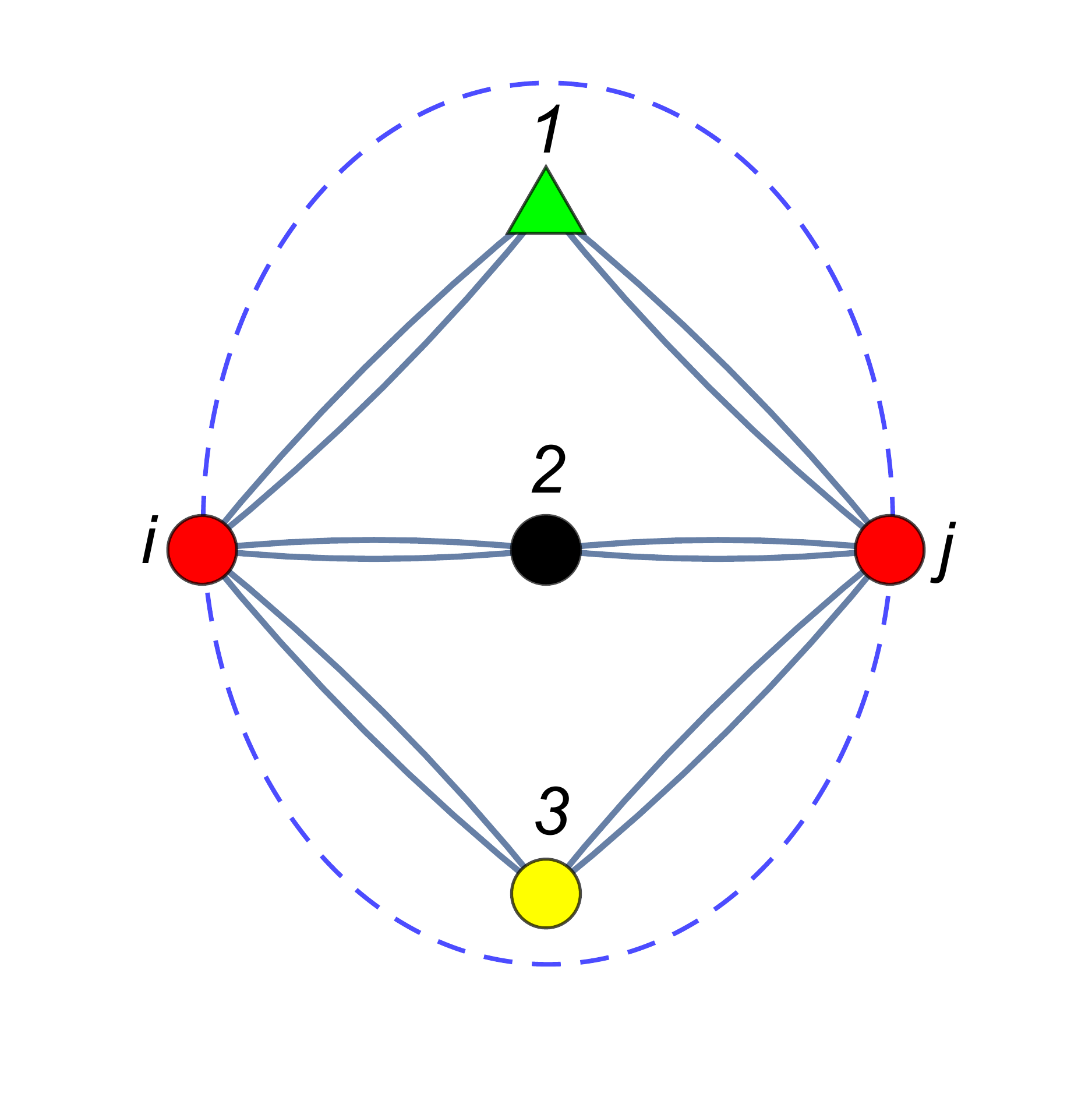}
\begin{center}
Figure 12: \,{\small {\rm CHY diagram for n=3 at one loop ({\rm 3-gon}).\,}}
\end{center}
\end{center}
Whose non-zero allowed cut-configurations are given by 
\begin{center}
\includegraphics[scale=0.2]{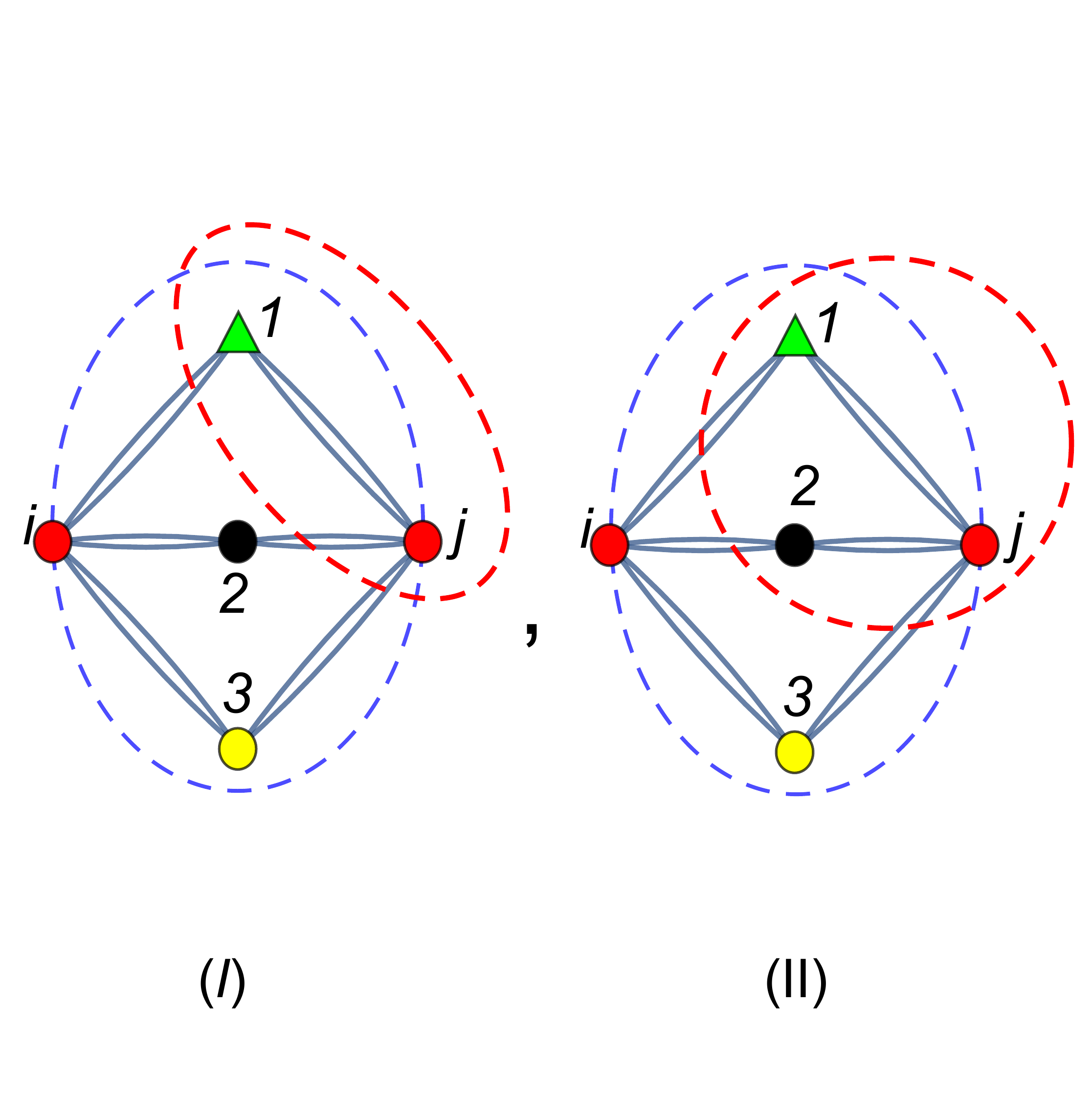} \, \, ,
\end{center}
\begin{center}
Figure 13: \,{\small {\rm All possible non-zero configurations for the  ${\rm 3-gon}$, up to $( i\,\leftrightarrow \, j)$.\,}}
\end{center}
plus the ones coming from the exchange $( i\,\leftrightarrow \, j)$.
Applying the $\Lambda-$algorithm as explained in the section above, we found the solution for each configuration
\begin{center}
\includegraphics[scale=0.32]{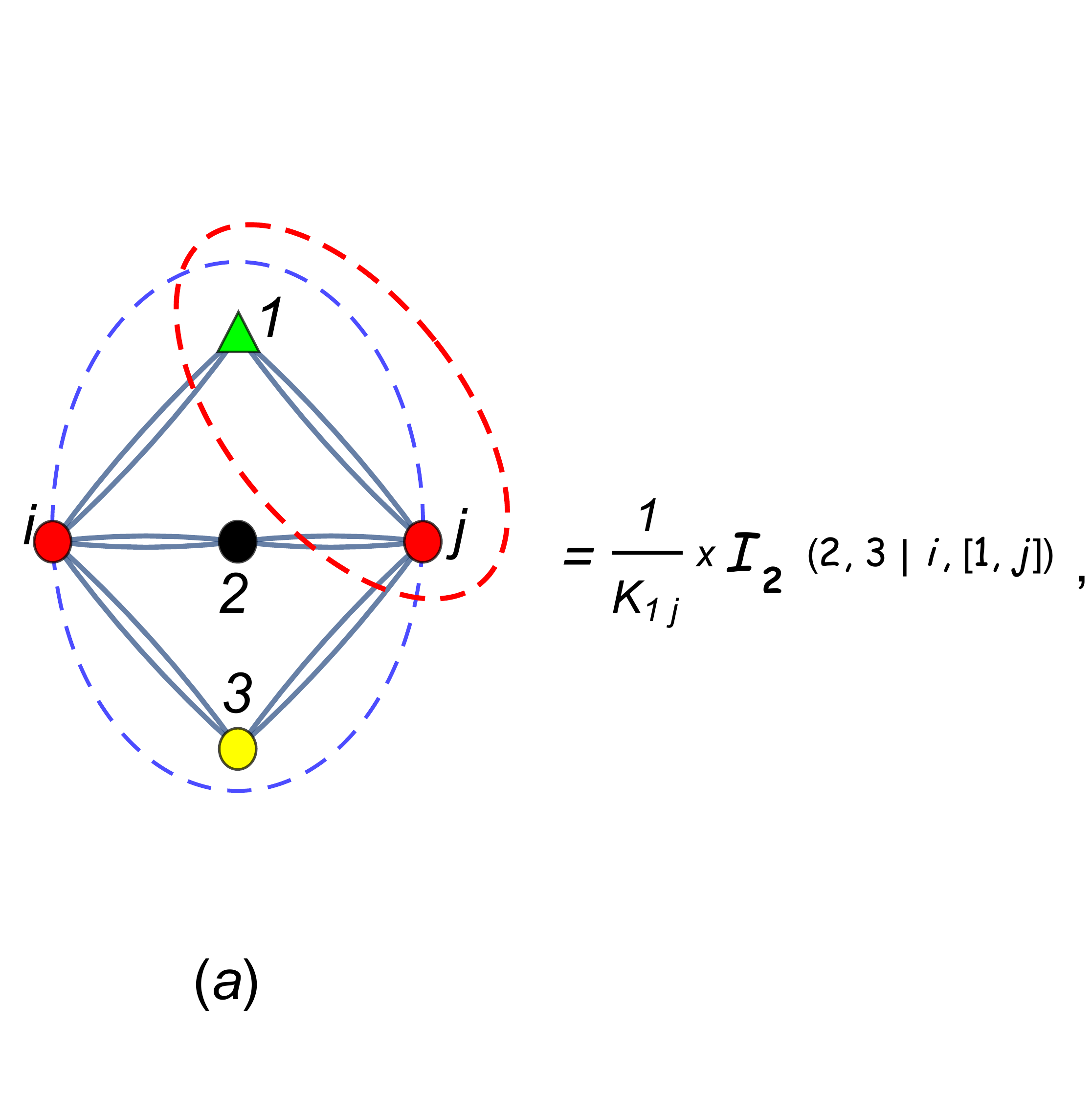}
\includegraphics[scale=0.32]{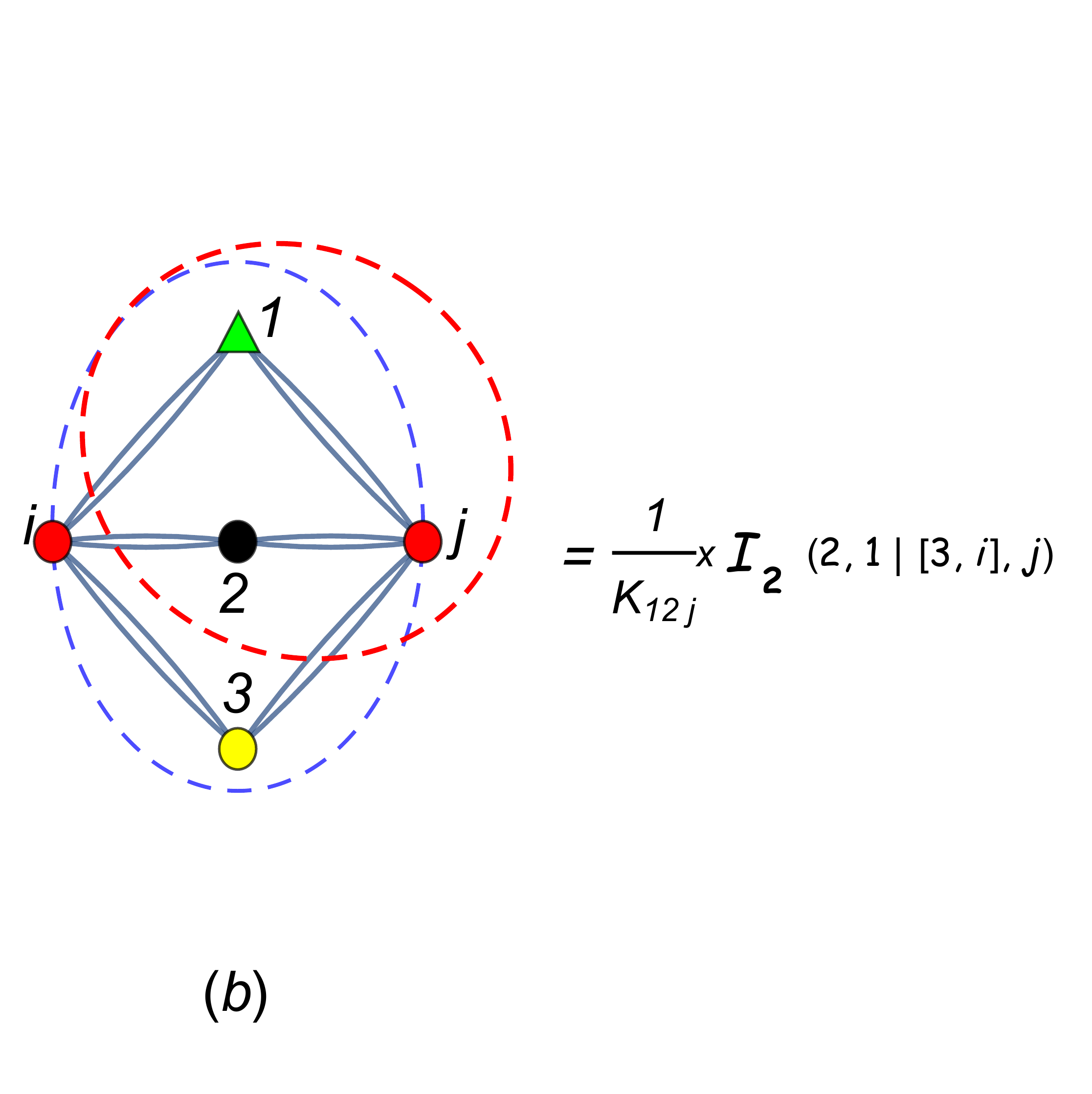}.
\end{center}
\begin{center}
Figure 14. \,{\small {\rm  Solution for the non-zero configurations for the ${\rm 3-gon}$, up to $( i\,\leftrightarrow \, j)$.}}
\end{center}
Replacing the solution given in figure 14,  one obtains (plus the exchange $( i\,\leftrightarrow \, j)$ )
\begin{align}\label{triangleOff} 
{\cal I}_3(1,2,3|i,j)=&{{\cal I}_2 (2,3|i,[1,j])\over k_{1j}}+{{\cal I}_2(2,1|[3,i],j)\over k_{12j}}\nonumber\\
& + {{\cal I}_2 (2,3 | [1,i],j)\over k_{1i}}+{{\cal I}_2(2,1| i, [3,j])\over k_{12i}}
\end{align}
Clearly, this expression agrees with the recurrence relation  in \eqref{Expan1}.
Using the building block given in \eqref{Gbubble} and replacing, $k_i^\mu=-k^\mu_j=\ell^\mu$, the above expression becomes  
\begin{align}\label{triangleOn}  
{\cal I}^{\rm 3-gon}_3(1,2,3) &= {\cal I}_3(1,2,3|\ell,-\ell)\\
&={-1\over\ell\cdot k_1}\left[{1\over\ell\cdot k_2}+{1\over -\ell\cdot k_2+k_{12}}\right]-{1\over \ell\cdot k_{1}+\ell\cdot k_{2} -k_{12}  }\left[{-1\over\ell\cdot k_2}+{1\over \ell\cdot k_2+k_{23}}\right]\,\nonumber\\
&+{1\over\ell\cdot k_1}\left[{-1\over\ell\cdot k_2}+{1\over \ell\cdot k_2+k_{12}}\right]+{1\over \ell\cdot k_{1}+\ell\cdot k_{2} + k_{12}  }\left[{1\over\ell\cdot k_2}+{1\over -\ell\cdot k_2+k_{23}}\right]\,\nonumber
\end{align}
This last expression vanish trivially whenever all the external momenta are on-shell, i.e. $k_{12}=k_{23}=k_{13}=0$ ,which is the expected result. 
 In order to compare with the partial fractions expansion given by
 \begin{equation}
 \sum_{\s\in S_3}\frac{1}{\ell\cdot k_{\s 1}\,(-\ell\cdot k_{\s 3})}=   
\frac{-2}{\ell\cdot k_1\, \ell\cdot k_3}+\frac{-2}{\ell\cdot k_2 \,  \ell\cdot k_3 }+\frac{-2}{\ell\cdot k_1\, \ell\cdot k_2} ,
\end{equation}  
let us consider one of the particles, namely $k_3^\mu$, as being off-shell, $k_3^2\neq 0$. After some algebra using momentum conservation, we manage to rewrite (\ref{triangleOn}) as,
\begin{align}
{\cal I}^{\rm 3-gon}_3(1,2,3) &={-2\over \ell\cdot k_1\,\ell\cdot k_2}+ {-2\over \ell\cdot k_2  (\ell\cdot k_3+\half k_3^2)}
+ {-2\over \ell\cdot k_1  (\ell\cdot k_3+\half k_3^2)}
\\
&+ k_3^2 \left(\frac{\ell\cdot k_3}{\ell\cdot k_1\,\ell\cdot k_2 (\ell\cdot k_3+\half k_3^2) (\ell\cdot k_3-\half k_3^2)}\
\right)\nonumber\, .
\end{align}
Therefore the expression coming from the $\Lambda$ expansion coincides with the one coming from Feynman diagrams when $k_3^2$ becomes on-shell.

\subsection{Four-particles scattering} 
Let us now to consider the next easiest scattering. The box Feynman diagram,
\begin{center}
\includegraphics[scale=0.55]{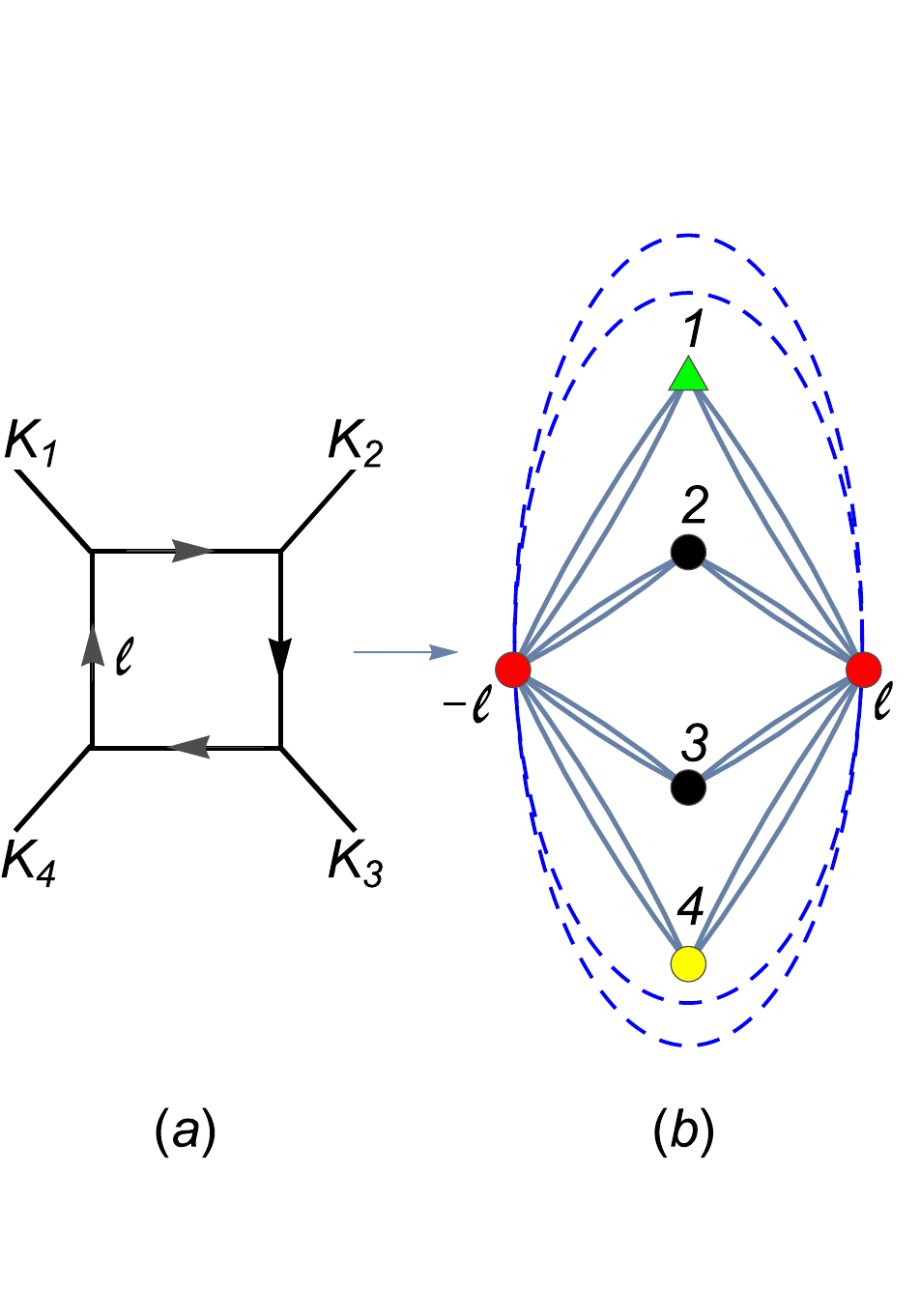}\,\, .
\begin{center}
Figure 15: \,{\small {\rm Box Feynman diagram (a)  and CHY graph (b).\,}}
\end{center}
\end{center}

After using (\ref{partialfrac}), this box Feynman diagram becomes
\be\label{a4feyn}
\ell ^2 \,\,{\cal I}_{\rm Feynman }^{\rm 4-gon}=\sum_{\sigma\in S_4}{1\over \ell\cdot k_{\sigma_1}(\ell\cdot(k_{\sigma_1}+k_{\sigma_2})+k_{\sigma_1}\cdot k_{\sigma_2})(-\ell\cdot k_{\sigma_4})}\,.
\ee 
On the other hand, by using $\Lambda$ algorithm we have the non zero allowable configurations
\begin{center}
\includegraphics[scale=0.55]{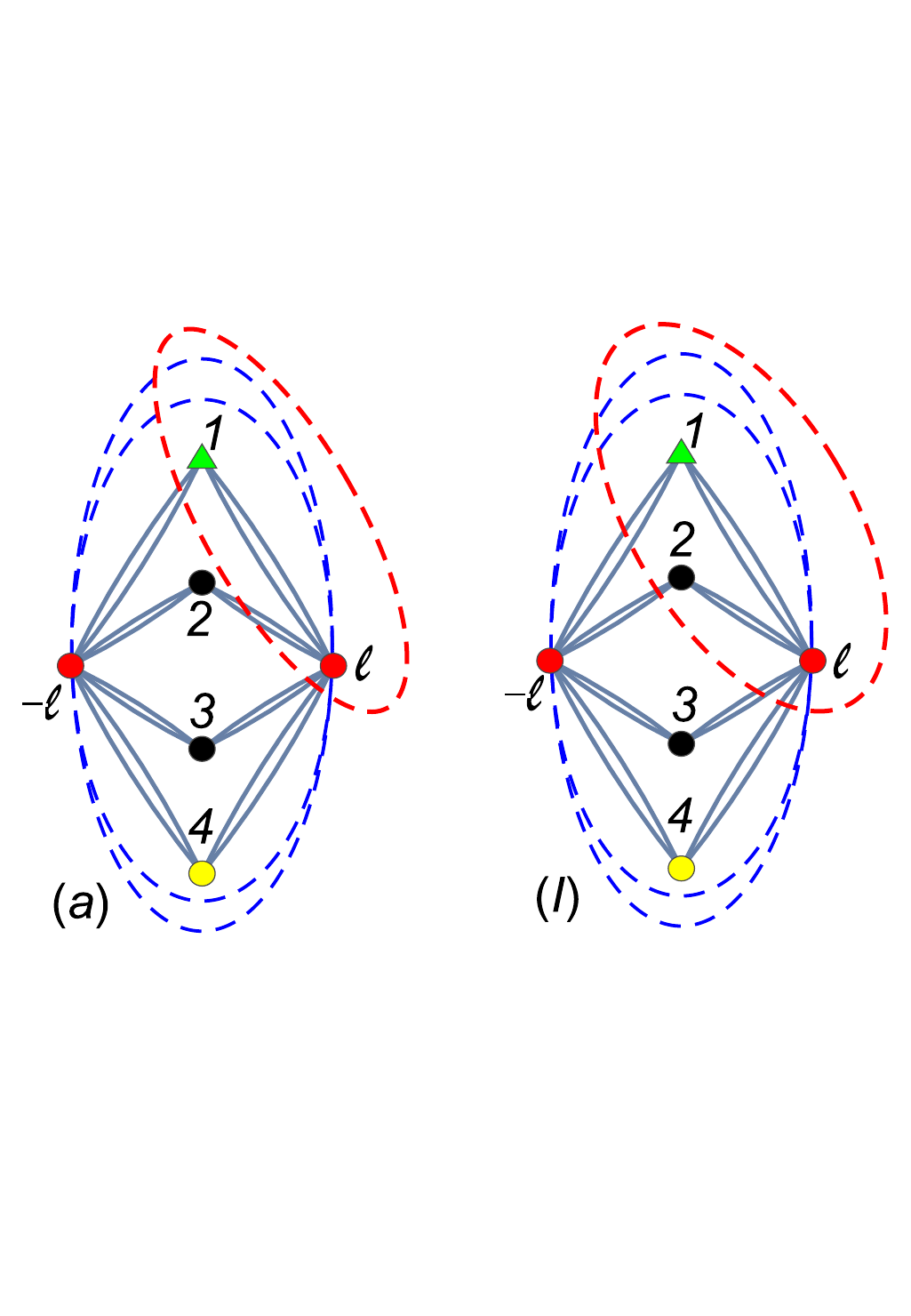}
\includegraphics[scale=0.55]{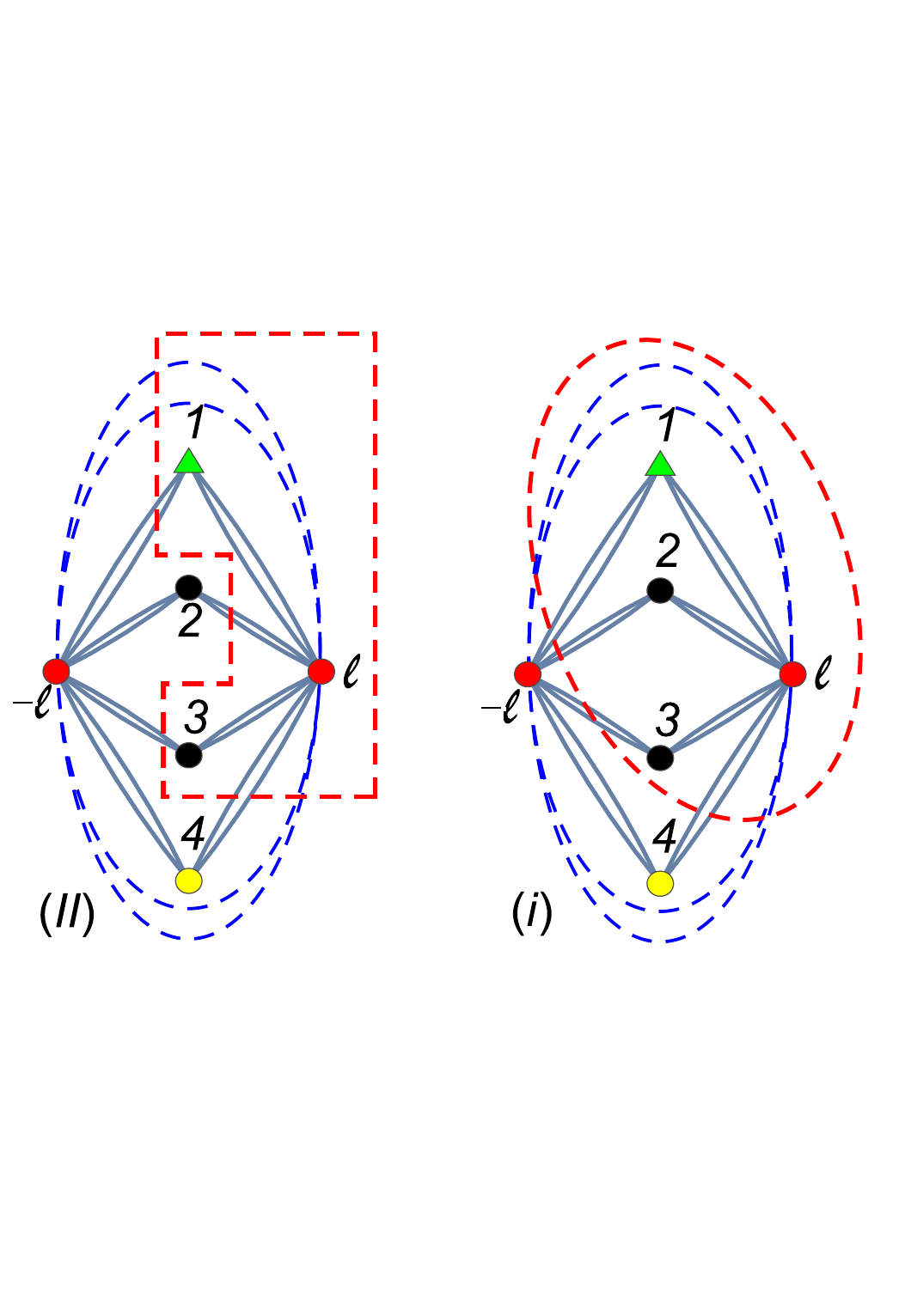}
\begin{center}
Figure 16: \,{\small {\rm $\Lambda$-algorithm for the ${\rm 4-gon}$, up to, $\ell \,\leftrightarrow \, -\ell$,  symmetry.\,}}
\end{center}
\end{center}
Computing this configurations we obtain
\begin{align}
{\cal I}_4^{\rm 4-gon}(1,2,3,4)& = \frac{{\cal I}_3(2,3,4|-\ell, [1,\ell])}{\ell\cdot k_1}+ \frac{  {\cal I}_2(2,1|[-\ell, 3,4  ],\ell) \,\, {\cal I}_2(3,4|-\ell, [1,2 ,\ell ])   }{\ell\cdot (k_1+k_2)+k_{12}}\\
&
+ \frac{ {\cal I}_2(3,1|[-\ell, 2,4  ],\ell) \,\, {\cal I}_2(2,4|-\ell, [1,3 ,\ell ])   }{\ell\cdot (k_1+k_3)+k_{13}}
+\frac{{\cal I}_3(2,3,1|[-\ell,4  ],\ell)   }{\ell\cdot (k_1+k_2+k_3)+k_{123}}\nonumber\\
&+ (\ell \,\leftrightarrow \, -\ell) \nonumber,
\end{align}
which is in agreement with the recurrence relation found in \eqref{Expangon}.
Written this expression in terms of the fundamental block, ${\cal I}_2(a,b|i,j)$ one obtains
\begin{align}\label{a4L}
&{\cal I}_4^{\rm 4-gon}(1,2,3,4)=\\
&{1\over\ell\cdot k_1}\left[{{\cal I}_2 (3,4\, |\,[1,2,\ell],-\ell)\over \ell\cdot k_2 +k_{12} }+{{\cal I}_2 (3,4,\,|\, [1,\ell] , [2,-\ell] )\over-\ell\cdot k_2} 
+{{\cal I}_2 (3,2\,|\, [1,\ell] , [4,-\ell] )\over\ell\cdot(k_2+k_3)}+{{\cal I}_2 (3,2\, |\, [1,4,\ell] ,-\ell)\over -\ell\cdot(k_2+k_3) + k_{23}}\right]\nonumber\\
&-{1\over\ell\cdot k_4}\left[{{\cal I}_2 (3,1\,|\, \ell , [2,4,-\ell] )\over -\ell\cdot k_{2}+k_{24}}+{{\cal I}_2 (3,1\, |\, [2,\ell] ,[4,-\ell])\over\ell\cdot k_2}
+{{\cal I}_2 (3,2\, |\, [1,\ell] \, ,[4,-\ell])\over-\ell\cdot(k_2+k_3)}+{{\cal I}_2 (3,2\, |\, \ell,\,[1,4,-\ell])\over\ell\cdot (k_2+k_3)+k_{23}}\right]\nonumber\\
&+{{\cal I}_2(2,1|[-\ell, 3,4  ],\ell) \,\, {\cal I}_2(3,4|-\ell, [1,2 ,\ell ])  \over  \ell\cdot (k_1+k_{2})+k_{12}}
+{   {\cal I}_2(3,1|[-\ell, 2,4  ],\ell) \,\, {\cal I}_2(2,4|-\ell, [1,3 ,\ell ]) 
\over \ell\cdot (k_1+k_3)+k_{13}}
+(\ell\, \leftrightarrow      \,-\ell) \nonumber
\end{align} 
By using momentum conservation one can check that \eqref{a4L} is exactly \eqref{a4feyn}.

\subsection{Five-particles scattering}

Let us display now the explicit expansion for the five-particles case,
\begin{center}
\includegraphics[scale=0.5]{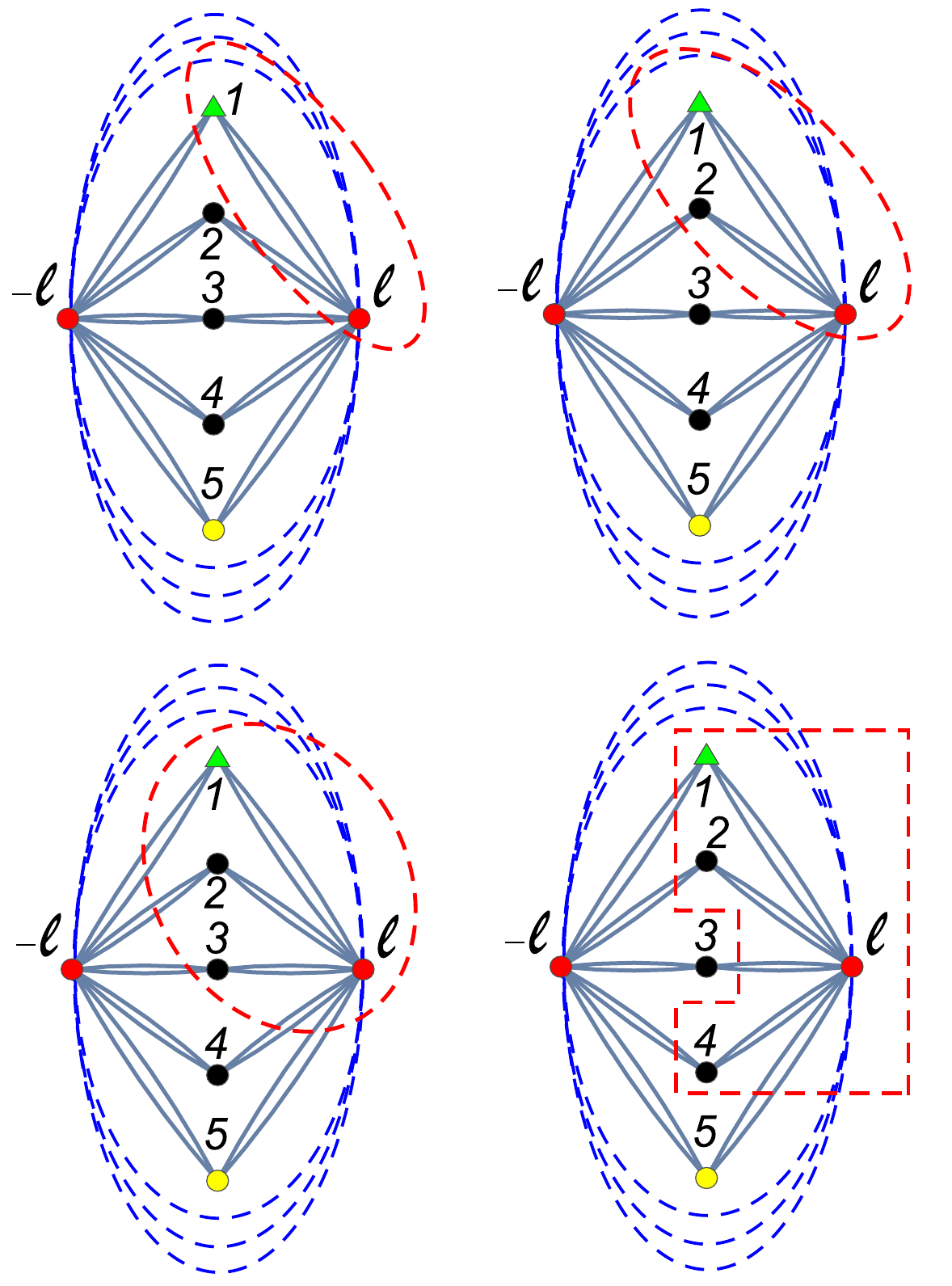}
\includegraphics[scale=0.5]{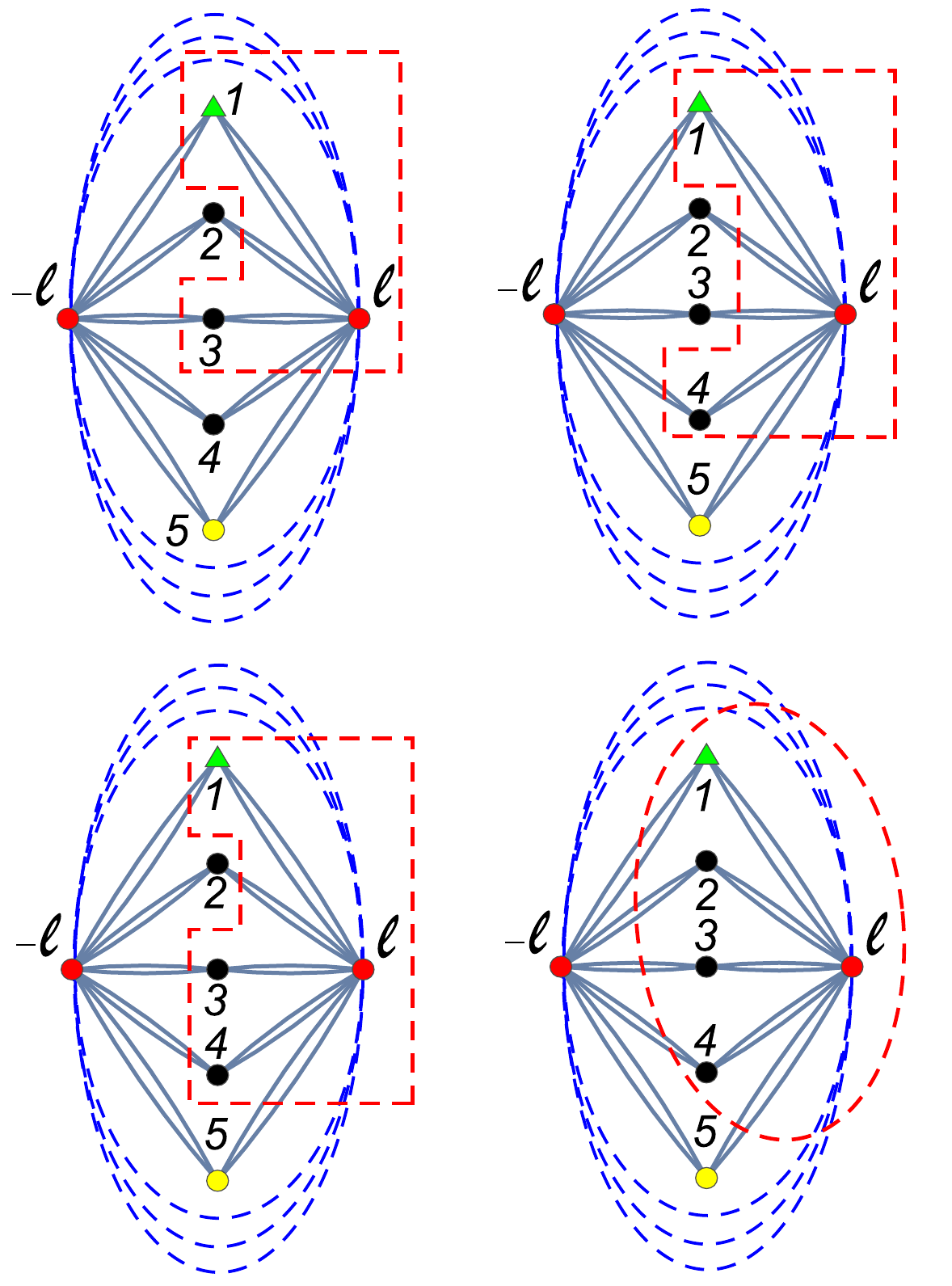}
\begin{center}
Figure 17: \,{\small {\rm $\Lambda$-algorithm for the ${\rm 5-gon}$, up to, $\ell \,\leftrightarrow \, -\ell$,  symmetry.\,}}
\end{center}
\end{center}

\begin{align}
&{\cal I}_5^{\rm 5-gon}(1,2,3,4,5)= \frac{{\cal I}_4(2,3,4,5|-\ell, [1,\ell])}{k_{1\ell}}+ \frac{{\cal I}_3(3,4,5|-\ell,[1,2,\ell]) \,\, {\cal I}_2(2,1|[3,4 ,5,-\ell],\ell)  }{k_{12\ell}}\nonumber\\
&
+ \frac{{\cal I}_3(2,4,5|-\ell,[1,3,\ell]) \,\, {\cal I}_2(3,1|[2,4,5,-\ell],\ell)   }{k_{13\ell}}
+\frac{{\cal I}_3(2,3,5|-\ell,[1,4,\ell]) \,\,{\cal I}_2(4,1|[2,3,5,-\ell],\ell)  }{k_{14\ell}}\nonumber\\
&
+ \frac{{\cal I}_2(4,5|-\ell,[1,2,3,\ell]) \,\, {\cal I}_3(2,3,1|[4,5,-\ell],\ell)   }{k_{123\ell}}
+\frac{{\cal I}_2(3,5|-\ell,[1,2,4,\ell]) \,\,{\cal I}_3(2,4,1|[3,5,-\ell],\ell)  }{k_{124\ell}}\nonumber\\
&
+ \frac{{\cal I}_2(2,5|-\ell,[1,3,4,\ell]) \,\, {\cal I}_3(3,4,1|[2,5,-\ell],\ell)   }{k_{134\ell}}
+\frac{{\cal I}_4(2,3,4,1|[5,-\ell],\ell)  }{k_{1234\ell}}\nonumber\\
&+ (\ell \,\leftrightarrow \, -\ell) \nonumber,
\end{align}
 which can be rewritten in terms of the basic piece (\ref{Gbubble}).   
We have checked numerically this result against the expected ${\rm  5-gon}$ result (\ref{pfe}).
For higher points it is equally straightforward to perform numerical checking.


\section{Discussion}\label{Sec9}

In this paper we have proposed a new approach called the elliptic scattering equations, which is a generalization of the $\L$ scattering equations prescription \cite{Gomez:2016bmv}. After integrate the modular parameter of the torus by using the global residue theorem, the amplitude splits in two regions connected by a fixed nodal point and a brach cut. This in turns, allows us to implement the $\Lambda-$algorithm, which provides us with a new recurrence relation expansion in terms of tree-level off-shell amplitudes for the ${\rm n-gon}$. This expansion is explicitly  read as
\begin{align}\label{ExpanDis}
&{\cal I}_{n}(1,2,\dots ,n | i,j)=\\
&\sum_{p=0}^{n-2}\sum_{s_p\in S(p)}
{{\cal I}_{n-p-1}(\hat{s}_{p},n\,|\, i,[1,s_{p},j])\times{\cal I}_{p+1}(s_{p},1\,  |\, [i,\hat{s}_{p},n],j)\over k_{1, s_{p}, j}}+(i\leftrightarrow j)\, ,\nonumber 
\end{align}
where the $i$ and $j$ particles are off-shell (for more details refer to section \ref{ngonR}). This expansion has two fundamental  properties, which are not manifests, but they can be deduced from the integrand in \eqref{HyperSmatrixForwardtauG}. The first one is the invariance by permutation of labels $1,.\ldots, n-1$, i.e.
\begin{equation}
{\cal I}_{n}(1,2,\dots ,n | i,j)  = \frac{1}{(n-1)!}\sum_{\s\in S_{n-1}}{\cal I}_{n} (\s_1,\s_2,\ldots ,\s_{n-1},n | i,j),
\end{equation}
where $S_{n-1}$ is the permutation group of $n-1$ elements. In  the particular case when, $k_i^\mu=-k_j^\mu:=\ell^\mu$, then \eqref{ExpanDis} becomes invariant over all n-indices
\begin{equation}
{\cal I}_{n}(1,2,\dots ,n | \ell,-\ell)  = \frac{1}{n !}\sum_{\s\in S_{n}}{\cal I}_{n} (\s_1,\s_2,\ldots ,\s_{n} | \ell,-\ell),
\end{equation}
and this is the second property. These two properties have been checked numerically.

This new recurrence relation has been checked analytically up to  $n=4$ and numerically up to $n=9$.  Our results agrees with some of the ones presented recently in \cite{Geyer:2015jch},  albeit it is important to stress that the methods used are different in nature.

A straightforward generalization of the methods presented in this paper consist in the use of higher order curves (hyperelliptic curves),  in order to  compute amplitudes at  higher loops. A further natural task would be to tackle the two-loop  amplitude for planar $\phi^3$ diagrams. A natural extension of CHY for dealing with higher loops for the cubic scalar theory has been developed recently by Bo Feng in \cite{Feng:2016nrf}. We expect to perform higher loop computations in the near future.

It also would be important to apply the elliptic curve formalism to  one-loop scattering amplitudes in other interesting theories, such as Yang-Mills, Supergravity, biadjoint scalar, among others. This have been studied previously in \cite{He:2015yua,Geyer:2015bja,Baadsgaard:2015hia}. We believe that once the integrand for the corresponding theory is guessed, the application of the techniques used in this paper apply straightforwardly to those cases.

It will be also interesting to consider the recent approaches to the scattering equations for generic number of particles \cite{Cardona:2015eba, Cardona:2015ouc, Dolan:2015iln,Cachazo:2015nwa,Huang:2015yka} (see also \cite{Kalousios:2015fya,Sogaard:2015dba}), in order to look for hidden mathematical structures in the scattering equations at one-loop as it has been done for tree-level.

Finally, notice that in this new prescription  we have not imposed any restriction for the momentum dimension. Let us remember that in other recent approaches \cite{Mason:2013sva,Berkovits:2013xba,Geyer:2015bja}, the dimension constraint is a consequence from the Modular invariance. Nevertheless, although we do not know how Modular transformations works into the elliptic scattering equations, we believe that for our prescription the Modular invariance is not a fundamental symmetry. It is due to the fact that we are describing a field theory amplitude instead of a String theory amplitude\footnote{We thank to F. Cachazo and N. Berkovits the discussion on this point.}, roughly speaking we do not have an $\alpha^\prime$  parameter in our approach.  As a final remark, imposing scale invariant in \eqref{HyperSmatrix} (Scattering amplitude prescription)  under the scale transformation, $(q^\mu,y_a,\s_a,\l_1,\l_2)\,\,  \rightarrow  \,\, (\kappa\,q^\mu, \kappa^3\,y_a,\kappa^2\,\s_a,\kappa^2\,\l_1,\kappa^2\,\l_2),~\kappa\in\mathbb{C}^*$, which coming from the scale invariance of the elliptic curve, the momentum dimension is restricted to be $D=10$.

\vspace{5mm}


\acknowledgments

\vspace{3mm}

\noindent
It is our pleasure to thank to F. Cachazo, B. Feng, N. Berkovits and R. Huang for useful comments and discussions. C.C. thanks to the High Energy Group of the Institute of Modern Physics and Department of Mathematics of Zhejiang University in Hangzhou city, People's Republic of China, for hospitality during the completion of this work. H.G.  would like to thank the hospitality of Perimeter Institute,  Universidade de S\~ao Paulo (USP)
and Universidad Santigo de Cali,  where this work was developed. The work of C.C. is supported in part by the National Center for Theoretical Science (NCTS), Taiwan, Republic of China. The work of  H.G.  is supported by CNPq  grant 403178/2014-2.


\bibliographystyle{JHEP}
\bibliography{mybib}
\end{document}